\documentclass[a4paper,aps,prb,twocolumn,amsmath,superscriptaddress]{revtex4-2}
\usepackage{graphicx}
\usepackage{color}
\usepackage{epstopdf}
\usepackage{xcolor}

\begin{document}

\title{Influence of local strain on the optical probing of a Ni$^{2+}$ spin in a charged self-assembled quantum dot}

\author{K. E. Polczynska}
\affiliation{Univ. Grenoble Alpes, CNRS, Grenoble INP, Institut N\'{e}el, 38000 Grenoble, France}
\affiliation{Univ. of Warsaw, Faculty of Physics, ul. Pasteura 5, 02-093 Warsaw, Poland}
\author{S. Karouaz}
\affiliation{Univ. Grenoble Alpes, CNRS, Grenoble INP, Institut N\'{e}el, 38000 Grenoble, France}
\author{W.~Pacuski}
\affiliation{Univ. of Warsaw, Faculty of Physics, ul. Pasteura 5, 02-093 Warsaw, Poland}
\author{L.~Besombes}\email{lucien.besombes@neel.cnrs.fr}
\affiliation{Univ. Grenoble Alpes, CNRS, Grenoble INP, Institut N\'{e}el, 38000 Grenoble, France}

\date{\today}

\begin{abstract}

This study explores the optical properties of quantum dots doped with a Ni$^{2+}$ ion that interacts with a charged exciton. Systematic magneto-optical analysis reveals that the strain distribution at the Ni$^{2+}$ site significantly influences its spin structure. In positively charged dots dominated by in-plane biaxial strain, the three spins states of the Ni$^{2+}$ (S$_z$=0, S$_z$=$\pm$1) can be observed and the magneto-optical spectra enables a local strain anisotropy to be determined. However, in most of the dots, lower-symmetry strain mixes all the Ni$^{2+}$ spin states, thereby increasing the number of observed optical transitions. In charged dots, we identify optical transitions that share a common excited state. They form a series of $\Lambda$ levels systems that can be individually addressed optically to determine the energy level structure. Magneto-optical measurements demonstrate that the hole-Ni$^{2+}$ exchange interaction is antiferromagnetic and considerably stronger than the electron-Ni$^{2+}$ interaction. A spin-effective model that incorporates local strain orientation can successfully reproduce key experimental results. Furthermore, we demonstrate that low-symmetry terms in the hole-Ni$^{2+}$ exchange interaction must be considered in order to accurately describe the emission spectra details in a magnetic field.

\end{abstract}

\maketitle

\section{Introduction}

The spin state of a carrier in a semiconductor quantum dot (QD) or of an individual impurity in a semiconductor host can act as a bit of quantum information with some possible applications in quantum sensing \cite{Degen2017}, quantum communication \cite{Knaut2024} or even quantum computing \cite{Veldhorst2014}. In direct bandgap semiconductors, spins of confined carriers can be optically controlled providing a natural spin-photon interface. This is also the case for some defect-localized spins, but depends on a particular level configuration and optical selection rules \cite{Higginbottom2022, Gruber1997, Pingault2017}. For impurities that do not exhibit specific spin-dependent optical transitions or are not optically active, it is possible to exploit the electrical or optical properties of the host semiconductor to interact with the localized spin. In particular, confined carriers in QDs can be exchange coupled with embedded magnetic elements. This exchange interaction with carriers can be exploited to control a more localized spin with longer relaxation and coherence times. In the case of an optically active QD, this provides an optical access to a strongly localized spin \cite{Besombes2004,Leger2005,Kudelski2007,Bhatta2007,Kobak2014,Besombes2012,Besombes2014,Lafuente2016}. This has been demonstrated for some transition-metal elements in II-VI and III-V semiconductors and could be extended to other non-optically active individual magnetic defects.

We analyze here the case of nickel (Ni) in II-VI semiconductors. Nickel is expected to be inserted in a II-VI compound as a Ni$^{2+}$ ion. It is then a 3$d^{8}$ element which carries an electronic spin S=1 and an orbital momentum L=3. In addition, all of the stable isotopes of Ni have no nuclear spins. The orbital momentum combined with the spin-orbit coupling should induce a large spin-strain coupling for the ion. The $S_z=\pm$1 spin states could be directly coupled by dynamical in-plane strain making Ni$^{2+}$ a promising qubit for spin-mechanical systems \cite{Besombes2019,Tiwari2020JAP}. A Ni$^{2+}$ spin could for instance be efficiently coupled to the strain field of a surface acoustic wave \cite{Tiwari2020JAP} for a mechanical driving of the $S_z=\pm1$ spin states or to a nano-mechanical oscillator to probe or control its position by coherent control of the localized spin \cite{Teissier2014}.

Inserting a Ni$^{2+}$ion into a QD should provide optical access to the spin states $S_z=\pm 1$ of the atom, as long as the exchange interaction with the confined carriers is strong enough. The 3$d$ electrons of a transition metal are exchange coupled to the electronic bands of the host semiconductor. The coupling with electrons at the center of the Brillouin zone arises from the standard exchange interaction and is ferromagnetic \cite{Furdyna1988}. It is weaker than the exchange interaction with the holes which arises from the hybridization of the 3$d$ orbitals of the magnetic atom and the $p$ orbitals of the host semiconductor, the so-called kinetic exchange. The resulting $p-d$ exchange interaction is strongly sensitive to the energy splitting between the 3$d$ levels and the top of the valence band and is usually anti-ferromagnetic \cite{Furdyna1988,Kacman2001,Blinowski1992}.

To demonstrate the optical access to a Ni$^{2+}$ spin, we performed the magneto-optic spectroscopy and resonant optical spectroscopy of Ni$^{2+}$-doped CdTe/ZnTe QDs. The studied QDs are in average p-doped and in the simplest case the positively charged exciton interacting with a Ni$^{2+}$ (X$^{+}$-Ni$^{2+}$) presents five photoluminescence (PL) lines at zero magnetic field. Under a longitudinal magnetic field, a characteristic cross-like behavior enables the local strain anisotropy to be determined. In most of the dots however, X$^{+}$-Ni$^{2+}$ shows nine lines corresponding to all possible transitions between the three spin states of Ni$^{2+}$ interacting with a single electron or hole in the excited or ground state of the QD, respectively. Resonant excitation reveals a series of $\Lambda$ levels that can be individually addressed optically. An analysis of the line energy spacing in the PL and resonant-PL spectra allows the splitting of the three Ni$^{2+}$ spin levels in the excited and ground states of the charged dot to be measured independently. The magneto-optic properties of X$^{+}$-Ni$^{2+}$ allow the identification of the main parameters controlling the interaction of the magnetic atom with the confined carriers. A comparison with a spin-effective model shows that the strain distribution at the position of the magnetic atom is crucial to the structure of the PL spectra. Additionally, an accurate description of the emission spectra requires consideration of low-symmetry terms in the hole-Ni$^{2+}$ exchange interaction arising from hole subbands mixing.

The rest of the paper is organized as follows: After a brief introduction of the sample and experiments in Sec. II, we describe in Sec. III the PL of positively charged Ni$^{2+}$-doped QDs. We detail in particular their magneto-optical properties and show that resonant PL measurements allow to unambiguously relate the optical transitions to the strain induced spin level structure. A spin-effective model is then discussed in Sec. V to identify the main parameters controlling the emission spectra of charged Ni$^{2+}$ doped QDs. The conclusion is given in Sec. VI.

\section{Experimental details}

We have studied Ni-doped self-assembled CdTe/ZnTe QDs grown by molecular beam epitaxy on a GaAs (100)-oriented substrate. During the CdTe deposition process a low level $\delta$ doping with Ni is introduced. As it has been shown for other magnetic elements, this can lead to some QDs containing a single magnetic atom \cite{Besombes2004,Kudelski2007,Kobak2014}. 

Single QDs are studied by optical micro-spectroscopy at liquid helium temperature (T= 4.2 K). The sample is mounted on x,y,z piezo actuators (Attocube, ANPx(z)101) and placed in a vacuum tube under a low pressure of He exchange gaz. The tube is immersed in the variable temperature insert of the cryostat filled with liquid He. The sample temperature is measured by a sensor located in the copper sample holder. The cryostat is equipped with a vectorial superconducting coil and a magnetic field of up to 9T along the QD growth axis and 2T in the QD plane can be applied.

The PL is generated using either a laser diode at 505 nm (Oxxius LBX-505) for non-resonant excitation or with a resonant tunable dye laser (Coherent CR599 with rhodamine 6G) and collected using a microscope objective with a high numerical aperture  (NA=0.85). For spectral analysis, the PL is dispersed by a two-meters double grating spectrometer (Jobin Yvon U1000 with 1800 gr/mm gratings giving a spectral resolution of around 25 $\mu$eV in the spectral range used) and detected by a cooled Si charge-coupled-device camera (Andor Newton). For the PL excitation (PLE) measurements, the power of the tunable dye laser is stabilized with an electro-optical variable attenuator (Cambridge Research $\&$ Instrumentation, LPC) when its wavelength is tuned. A half-wave plate in front of a linear polarizer is used to analyze the linear polarization of the PL. For circular polarization measurements, a quarter-wave plate oriented at 45° to the linear polarization direction of detection is inserted in the detection path. When inserted in both the excitation and detection paths, it also permits to select co- or cross-circularly polarized PL.

\begin{figure}[hbt]
\centering
\includegraphics[width=1\linewidth]{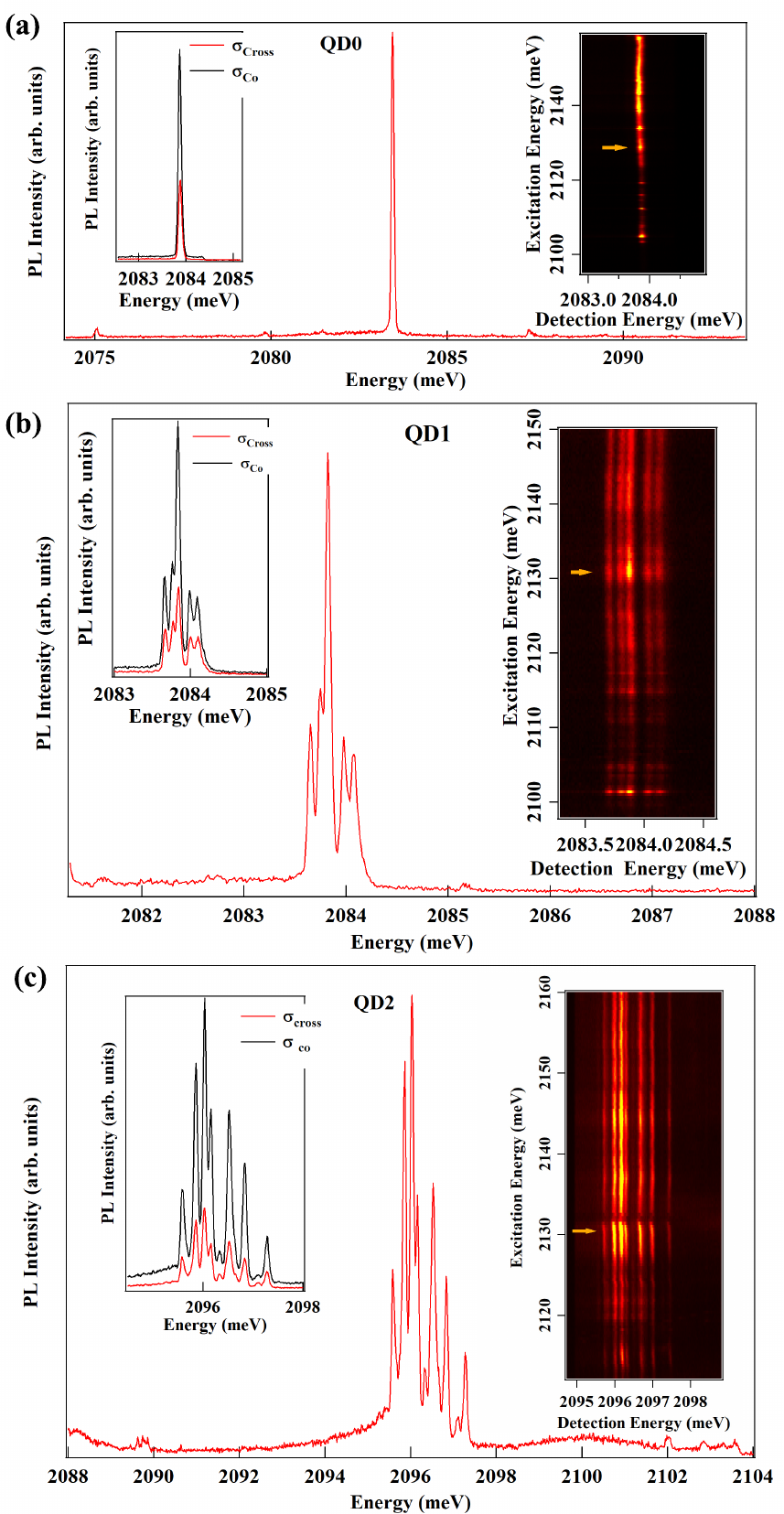}
\caption{(a) PL of X$^+$ in a non-magnetic QD (QD0). (b) and (c) PL of X$^{+}$ in two Ni$^{2+}$-doped QDs, QD1 and QD2 respectively. Right insets: corresponding PLE intensity map. Left insets: Co and cross circularity polarized PL spectra for an excitation energy pointed by the arrow on the PLE maps.}
\label{Fig:iso}
\end{figure}

\section{Optical properties of positively charged Ni$^{2+}$-doped quantum dots.}

In the studied Ni-doped sample, the QDs are on average positively charged. The background p-doping originates from the inherent p-type conductivity of ZnTe. The incorporation of Ni ions in ZnTe can also enhances its p-type character as Ni is a deep acceptor in ZnTe and could be observed in both its Ni$^{2+}$ and Ni$^{+}$ oxidation states \cite{Buktiar2024,Kaufmann1984}.

\begin{figure}[hbt]
\centering
\includegraphics[width=1\linewidth]{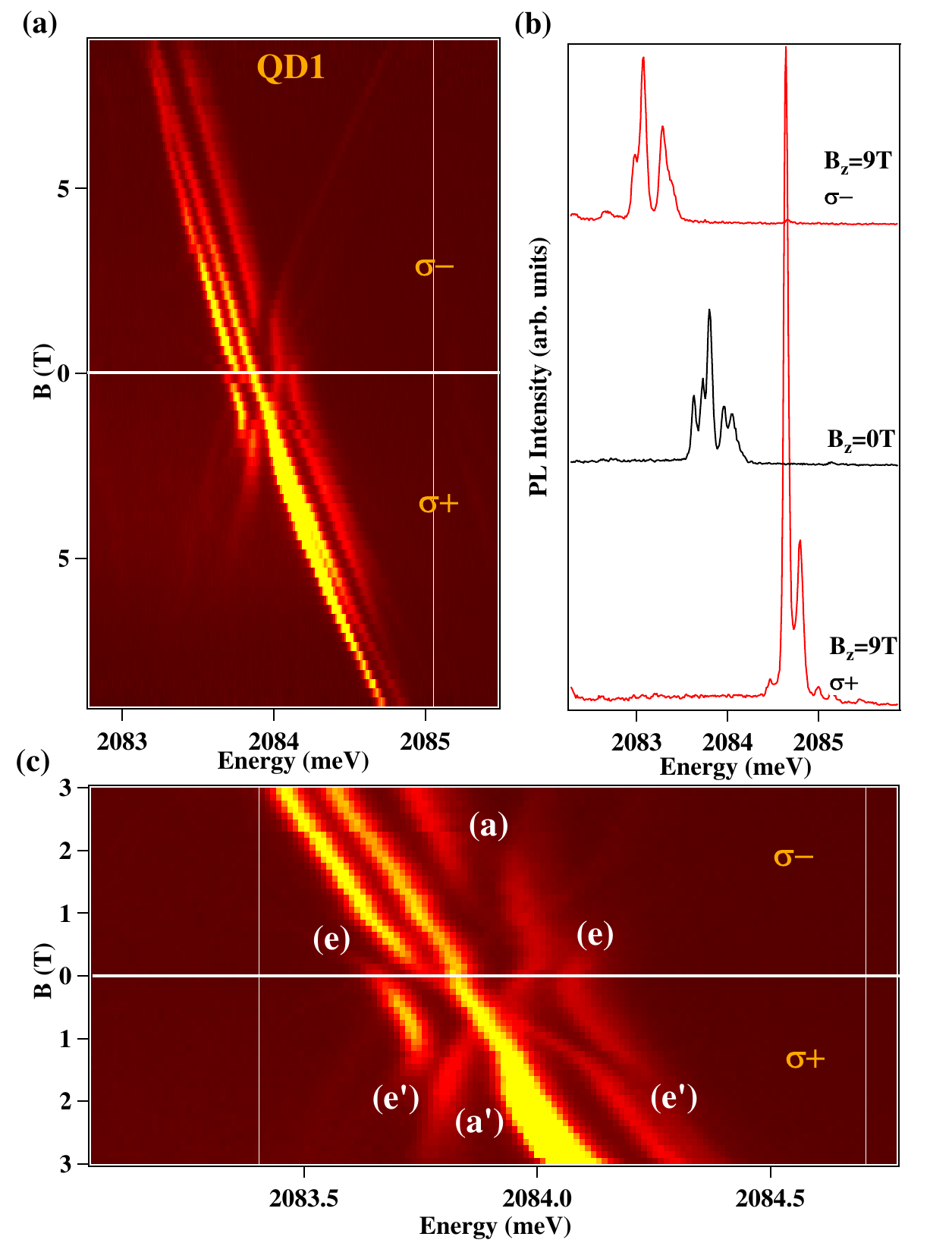}
\caption{(a) Longitudinal magnetic field (B$_z$) dependence of the emission of X$^{+}$-Ni$^{2+}$ in QD1. (b) PL spectra at zero field and under B$_z$=9T. (c) Details of the magneto-optic spectra at low magnetic field.}\label{Fig:isoB}
\end{figure}

Many of the observed optical transitions correspond to X$^+$ that can be identified by (i) the absence of fine structure splitting (in non-magnetic QDs, the PL of X$^+$ simply consists in a single line) and (ii)  partially co-circular polarization under optical excitation within the energy range of the excited states of the dots \cite{Varghese2014} (see QD0 in Fig.~\ref{Fig:iso}(a)). Conversely, a negative circular polarization is observed for the negatively charged exciton in CdTe/ZnTe QDs together with a specific triplet state signature of the charged exciton in the PLE spectra \cite{Ware2005,LeGall2012,Besombes2023}. 

We analyze here the magneto-optical properties of X$^{+}$-Ni$^{2+}$. In the excited state of a p-doped QD ({\it i.e.} X$^+$), the exchange interaction of the two holes with the spin of the Ni$^{2+}$ ion can be neglected. X$^+$-Ni$^{2+}$ can be approximated to an electron coupled to the Ni$^{2+}$ spin. In the ground state, after the recombination of the electron-hole (e-h) pair, the Ni$^{2+}$ spin interacts with the resident hole spin. For a spin S=1, three emission lines are expected for X$^+$-Ni$^{2+}$: a central line associated with S$_z$=0 and two outer lines associated with S$_z=\pm$1.

\begin{figure*}[!hbt]
\centering
\includegraphics[width=1\linewidth]{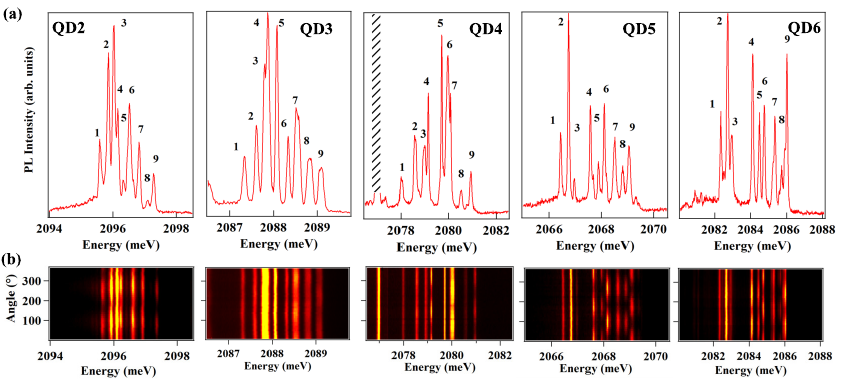}
\caption{(a) PL of charged excitons in five Ni$^{2+}$-doped QDs shown in increasing order of their total energy splitting. (b) Corresponding linear polarization PL intensity maps. The direction of polarization is measured with respect to the [100] or [010] axis of the sample (i.e., at 45~° from the easy cleavage axis of the substrate).}
\label{Fig:NiCPL}
\end{figure*}

\subsection{Magneto-optical properties of X$^+$-Ni$^{2+}$}

A wide variety of spectra are indeed observed in charged and Ni$^{2+}$-doped dots. The simplest case (QD1) is shown in Fig.~\ref{Fig:iso}(b). It consists of five lines, a central line with a doublet on each side. All lines are excited simultaneously by a resonant laser tuned on an exited state of the dot and the emission is mainly co-polarised with the excitation laser as expected for X$^+$ (Inset of Fig.~\ref{Fig:iso}(b)). 

Under a longitudinal magnetic field (B$_z$), the Zeeman effect causes each line to split, resulting in a distinctive PL intensity map (Fig.~\ref{Fig:isoB}). At low field, a cross-like pattern emerges, appearing in a positive magnetic field in $\sigma+$ polarization. The dependence on B$_z$ suggests that the doublets on the outer lines originate from anti-crossings, which also occur around B$_z$ $\approx$1.2T.

\begin{figure*}[!hbt]
\centering
\includegraphics[width=1\linewidth]{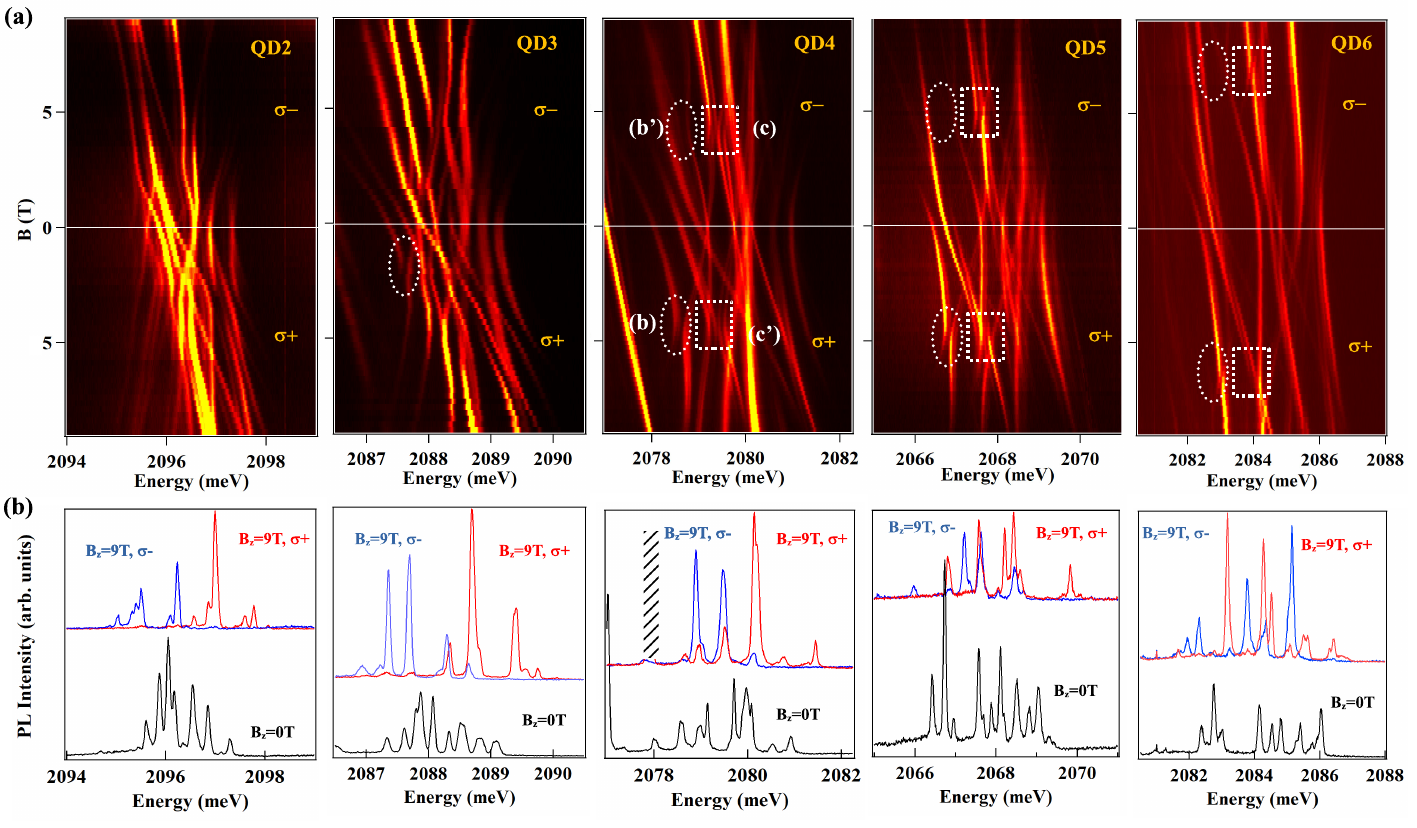}
\caption{(a) Intensity map of the longitudinal magnetic field dependence of the PL of five charged Ni$^{2+}$-doped QDs (see Fig.~\ref{Fig:modB}(c) for a detailed view of anticrossings (b), (b$^{\prime}$), (c), (c$^{\prime}$)). (b) Corresponding PL spectra at B$_z$=0T and circularly polarized PL spectra at B$_z$=9T. }
\label{Fig:mapB}
\end{figure*}

However, as displayed in Fig.~\ref{Fig:iso}(c), X$^{+}$-Ni$^{2+}$ usually presents a more complex emission spectra. The emission is still co-polarized with the excitation but the spectra deviate significantly from the simple structure of three lines expected for a spin S=1. The PL of 5 Ni$^{2+}$-doped and charged QDs are presented in Fig.~\ref{Fig:NiCPL}. Up to nine emission lines can be observed at zero magnetic field. They are labeled from (1) to (9) from the low to the high energy side of the spectra in Fig.~\ref{Fig:NiCPL}. The overall splitting of the spectra changes from dot to dot and some of the lines are partially linearly polarized (Fig.~\ref{Fig:NiCPL}(b)). The linearly polarized structure is particularly pronounced in QD5 and QD6, the dots with the largest zero-field splitting and appears predominantly on the high energy side of the spectra.

Magneto-optical measurements allow to extract more information about the carrier/magnetic atom interaction. The longitudinal magnetic field dependence of 5 charged and Ni$^{2+}$-doped QDs is presented in Fig.~\ref{Fig:mapB}. The 5 dots evolve in a similar way under B$_z$ with in particular a reduction in the number of lines observed at high field, where spectra are dominated by 2 or 3 lines in each circular polarization. This reduction reflects a thermalization on the spin states of the magnetic atom split by the applied magnetic field. 

Characteristic anti-crossings can also be seen in the PL intensity maps. This is clearer in dots with the largest splittings such as QD4, QD5, and QD6. For example in QD4, a large anti-crossing is observed on the lower energy line in $\sigma+$ polarization around B$_z$=5T (labelled (b) in Fig.~\ref{Fig:mapB}). Another anti-crossing appears in the center of the spectra, in $\sigma-$ polarization around B$_z$=5T (labeled (c) in Fig.~\ref{Fig:mapB}). Similar anti-crossings can sometimes be seen at the same energy and magnetic field on weaker intensity lines in the opposite circular polarization. They are labeled (b$^{\prime}$) and (c$^{\prime}$) in Fig.~\ref{Fig:mapB}. It can also be noticed in some of the dots that narrow lines in the center of the spectra split into a doublet under a weak B$_z$. This is particularly clear in QD4 where these lines are involved in the anti-crossing (c) and (c$\prime$) (see Fig.~(\ref{Fig:mapB}) and Fig.~\ref{Fig:modB}(c) for a more detailed view of the anti-crossings).

\subsection{Resonant fluorescence of X$^+$-Ni$^{2+}$.}

\begin{figure}[hbt]
\centering
\includegraphics[width=1\linewidth]{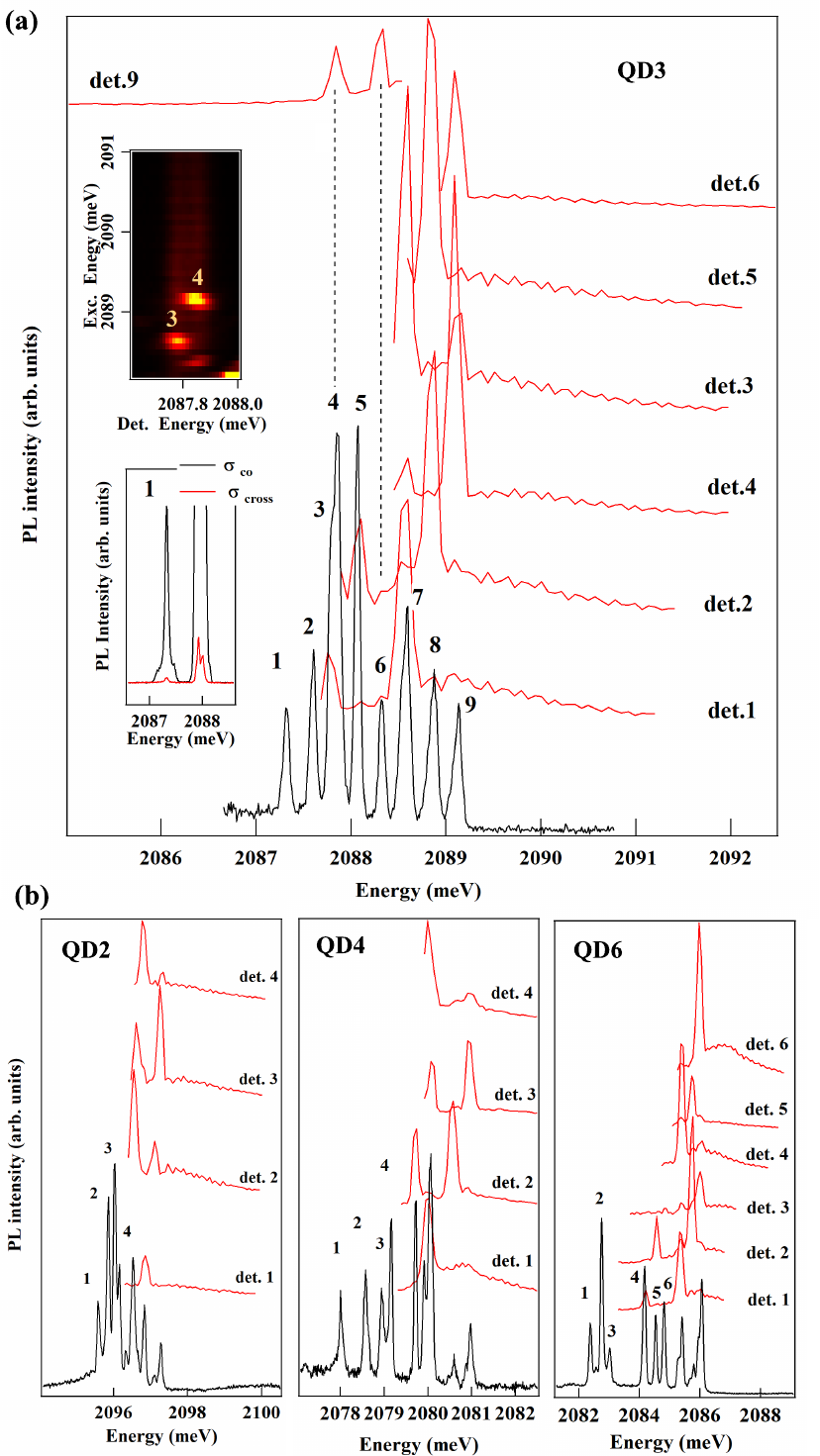}
\caption{(a) Non-resonant PL and PLE in QD3. PLE intensities are detected on lines (1), (2), (3), (4), (5) and (6) of X$^+$-Ni$^{2+}$ for an excitation on the high energy side and for a detection on the high energy line (9) and an excitation on the low energy side of the spectrum. Top inset: Example of resonant PL intensity map for a detection on lines (3) and (4) and an excitation on the high energy side. Bottom inset: Co and cross circularly polarized resonant PL spectra for a detection on (1) and an excitation on (7). (b) Similar PLE measurements on QD2, QD4 and QD6.}
\label{Fig:PLE}
\end{figure}

In order to gain more insight into the energy level structure of charged Ni$^{2+}$-doped QDs, we performed resonant excitation experiments on X$^+$-Ni$^{2+}$. The PL intensities detected on the low energy lines in QD3 when excited on the high energy lines in a cross-linearly polarized excitation/detection configuration are presented in Fig.~\ref{Fig:PLE}(a). In addition to a PL intensity background that decreases slightly with increasing the excitation energy, characteristic absorption resonances are observed.

The resonant PL background, also observed in non-magnetic QDs, arises from the absorption in the acoustic phonon side-band of the probed line \cite{Thomas2021}. Absorption resonances, on the other hand, result from an excitation transfer between the X$^+$-Ni$^{2+}$ levels and reveal some specific links between the spin states. The structure of the resonances is particularly clear in QDs with a large zero-field energy splitting such as QD3. For a detection on the low energy line (1) a large absorption is observed on line (7) and a weaker resonance on line (3). Line (2) is mainly excited by a laser tuned to line (8) with a weaker contribution for an excitation on (5). For a detection on line (4), a maximum of transfer is observed for an excitation on (9). A similar behavior is observed when successively detecting on lines (3), (5) and (6) where a transfer is observed from lines (7), (8) and (9) respectively. For a circularly polarized excitation on one of these absorption resonances, the PL is mainly co-polarized with the excitation (see inset of Fig.~\ref{Fig:PLE} (a) for QD3). This confirms the good conservation of the spin of the electron during the lifetime of X$^+$ coupled to the spin of the magnetic atom. It can also be noted that transfer occurs for an excitation on the low energy side of the spectrum and a detection on the high energy side. This is illustrated by the top curve in  Fig.~\ref{Fig:PLE}(a) where for a detection on the high energy line (9), absorption resonances are observed on the lowest energy lines (4) and (6).

Such resonant excitation experiments were systematically performed on many dots. Similar structures in the excitation spectra were observed, for example, for X$^+$-Ni$^{2+}$ in QD1 (small overall splitting), QD4 and QD6 (large overall splitting). The structure is clear in dots like QD6 which presents well separated 3-lines groups in the PL. Despite some broadening of the lines, a transfer sequence can be identified: For a detection on the low energy line (1) a large absorption is observed on line (7). Line (2) is mainly excited by a resonant laser on line (8) and line (3) by a transfer from line (9). A similar behavior is observed when successively detecting on lines (4), (5) and (6) where a transfer is observed from lines (7), (8) and (9) respectively. The energy positions and relative intensities of the absorption lines are similar in QD1, although the lower energy splitting of this dot makes it more difficult to precisely identify the corresponding PL lines. QD4 corresponds to an intermediate situation with a large overall splitting but no obvious 3-lines groups in the PL spectra. The observed structure of the excitation transfer sequence in this dot is very similar to QD3. 

Despite the PL spectra being quite different, the observed structure of the excitation transfer is very similar for all the dots. This suggests that the PL spectra can be organised into three groups of three lines. This is particularly evident in dots such as QD6, in which excitation transfer occurs between the low-, central, and high-energy lines of each group, respectively.

\section{Spin effective model of a positively charged Ni$^{2+}$-doped quantum dot.}

A large variety of spectra is observed in Ni$^{2+}$-doped QDs with a significant deviation from the simple structure of 3 lines expected for a spin S=1 interacting with a charged exciton. A detailed description of the magnetic atom spin structure and of the carriers-Ni$^{2+}$ exchange interaction is required to explain these spectra.

\subsection{A Ni$^{2+}$ ion in a strained II-VI semiconductor.}

Ni$^{2+}$ ion carries an electronic spin S=1 and an orbital momentum L=3.
According to Tanabe-Sugano diagrams, the fundamental $^{3}$F level (L=3, degeneracy 7) of a 3$d^{8}$ ion splits in T$_{d}$ symmetry into two triplets ($^{3}$T$_{1}$, $^{3}$T$_{2}$) and a singlet ($^{3}$A$_{2}$) at high energy. The fundamental triplet $^{3}$T$_{1}$ comes from a combination of orbitals $d_{xy}$, $d_{xz}$, $d_{yz}$. A reduction of the symmetry lifts the degeneracy of this triplet. For a symmetry reduction from T$_{d}$ to D$_{2d}$, the triplet $^{3}$T$_{1}$ splits into two distinct levels: a doublet $^{3}$E and a singlet $^{3}$B$_2$. If the symmetry reduction is due to an elongation along z (biaxial compression in the dot plane), the crystal field increases the energy of z-oriented orbitals ($d_{xz}$, $d_{yz}$) and decreases the energy of orbitals pointing in the xy plane. The fundamental is then the orbital singlet $^{3}$B$_2$ originating from the $d_{xy}$ orbital \cite{Smolenski2016}. Note that in the absence of strain or for low strain, the degeneracy of the ground orbital triplet is also lifted at low temperature by a static Jahn-Teller distortion \cite{Brousseau1988,Ando2002}. This ensures that the ground state is an orbital singlet.

\begin{figure}[hbt]
\centering
\includegraphics[width=1\linewidth]{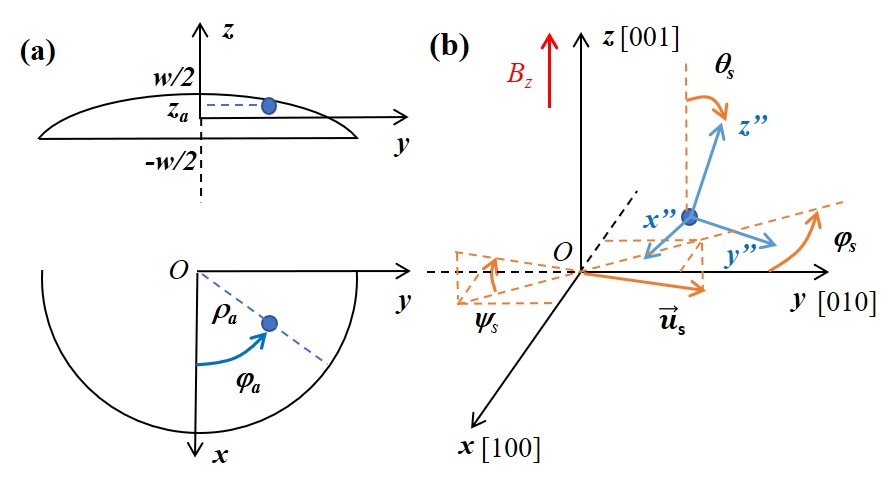}
\caption{QD geometry considered in the model illustrating in (a) the position of the magnetic atom, $\vec{r_a}(\rho_a,\varphi_a,z_a)$. (b) Orientation of the strain at the location of the magnetic atom. The strain frame ($x^{\prime\prime}$,$y^{\prime\prime}$,$z^{\prime\prime}$) presented in (b) is rotated by $\theta_s$ around $\vec{u_s}$. The presented configuration corresponds to $\theta_s>$0, $\varphi_s>$0 and $\psi_s<$0.}
\label{FigPosition}
\end{figure}

Taking also into account the spin-orbit coupling in this reduced symmetry induces a zero field spin splitting which depends on the local strain. In a self-assembled QD it is often assumed that the strain is biaxial with an elongation axis along the growth direction $z$. This results in a magnetic anisotropy described by the Hamiltonian D$_0$S$_z^{2}$ which is often sufficient to explain the main properties of most of the magnetic QDs \cite{Lafuente2016,Besombes2024}. A possible disorientation of the strain axis with respect to the [001] growth direction can however be taken into account and a general form of spin effective Hamiltonian is \cite{Krebs2009}:

\begin{eqnarray}
\mathcal{H}_{ZFS}=\frac{2}{3}D_{0}\left[S_{z^{\prime \prime}}^2-\frac{1}{2}\left(S_{x^{\prime \prime}}^2+S_{y^{\prime \prime}}^2\right)\right]+\frac{E}{2}\left(S_{x^{\prime \prime}}^2-S_{y^{\prime \prime}}^2\right) 
\end{eqnarray}

\noindent ($x^{\prime \prime},y^{\prime \prime},z^{\prime \prime}$) is the local frame of the strain at the position of the magnetic atom, with a principal axis along $z^{\prime \prime}$ . It is obtained by rotating the crystal frame by an angle $\theta_s$ around the unitary vector $\vec{u_s}$ (see Fig.~\ref{FigPosition}):

\begin{eqnarray}
\vec{u}_s\begin{pmatrix}
-sin(\varphi_s)\\ 
cos(\varphi_s)\\
sin(\psi_s)\end{pmatrix}/\sqrt{sin^2(\varphi_s)+cos^2(\varphi_s)+sin^2(\psi_s)}
\end{eqnarray}

Angular momentum operators S$_{\alpha^{\prime \prime}}$ are linked to S$_{\alpha}$ by the passive rotation transforming (x,y,z) into ($x^{\prime \prime},y^{\prime \prime},z^{\prime \prime}$). For strain axes coincident with the (x,y,z) QD frame, $\mathcal{H}_{ZFS}$ splits S$_z$=0 and S$_z$=$\pm1$ by D$_0$. Anisotropy of strain in the (x,y) plane mixes the $S_z=\pm1$ states, inducing a doublet split by $E$ \cite{Tiwari2020JAP}.

\subsection{X$^+$-Ni$^{2+}$ energy levels}

X$^+$-Ni$^{2+}$, which is composed of 2 holes, 1 electron and 1 Ni$^{2+}$ can be approximated as an electron interacting with the atom's spin. Its energy levels are described by the Hamiltonian:

\begin{eqnarray}
\mathcal{H}_{X^+,Ni}=E_g+2E_{hh}+\mathcal{H}_{e-Ni}+\mathcal{H}_{X^+-B}+\mathcal{H}_{Ni}
\label{Hamiltone}
\end{eqnarray}

\noindent where E$_g$ is the energy of the electron and E$_{hh}$ the energy of a heavy-hole (hh). E$_{hh}$=0 in the absence of heavy-hole/light-hole (hh/lh) mixing. $\mathcal{H}_{e-Ni}$ is the exchange coupling of the electron spin ($\vec{\sigma_e}$) and Ni$^{2+}$ spin ($\vec{S}$). A Heisenberg-type interaction is assumed and $\mathcal{H}_{e-Ni}=I_{eNi}\vec{S}\cdot\vec{\sigma_e}$ with $I_{eNi}$ the electron-Ni$^{2+}$ exchange integral. This interaction is ferromagnetic and usually weaker than the exchange interaction with the hole \cite{Furdyna1988}. A magnetic field couples to the electron spin via the Zeeman term and a diamagnetic shift of X$^+$ can be included resulting in 
$\mathcal{H}_{X^+-B}=g_e\mu_B\vec{B}\cdot\vec{\sigma_e}+\gamma B^2$ with g$_e$ the electron Lande factor and $\gamma$ a diamagnetic coefficient. $\mathcal{H}_{Ni}$ describes the fine structure of the Ni$^{2+}$ spin and its dependence on local strain and magnetic field: 

\begin{eqnarray}
\mathcal{H}_{Ni}=g_{Ni}\mu_B\vec{B}\cdot\vec{S}+\mathcal{H}_{ZFS}
\end{eqnarray}

\noindent where g$_{Ni}$ is the Ni$^{2+}$ Lande factors \cite{Silva2020}. At zero magnetic field, X$^+$-Ni$^{2+}$ is mainly controlled by $\mathcal{H}_{Ni}$ and the resulting energy levels are eventually influenced by a weak exchange interaction with the electron spin.

\subsection{Hole-Ni$^{2+}$ energy levels}

In the ground state of a charged dot, the hole-Ni$^{2+}$ complex is described by:

\begin{eqnarray}
\mathcal{H}_{h,Ni}=\mathcal{H}_{Ni}+\mathcal{H}_{h-Ni}+\mathcal{H}_{h-B}+\mathcal{H}_{VB}
\label{Hamiltonh}
\end{eqnarray}

\noindent where the levels are split at zero field by $\mathcal{H}_{Ni}$ and by the hole-Ni$^{2+}$  exchange interaction, $\mathcal{H}_{h-Ni}$. A magnetic field couples to the hole spin ($\vec{J_h}$) via the Zeeman term $\mathcal{H}_{h-B}=g_h\mu_B\vec{B}\cdot\vec{J_h}$ with g$_h$ a hole Lande factor which can depend on the direction of the magnetic field. It also couples to the Ni$^{2+}$ spin with an isotropic g$_{Ni}$ included in $\mathcal{H}_{Ni}$.

In the usual spherical symmetry approximation, $\mathcal{H}_{h-Ni}$ is described by the isotropic Heisenberg Hamiltonian

\begin{eqnarray}
\mathcal{H}_{h-Ni}=I_{hNi}\vec{J_h}\cdot\vec{S}
\label{ExIso}
\end{eqnarray}

\noindent where the hole-Ni$^{2+}$ exchange integral, 
$I_{hNi}\propto\beta\vert\Psi^{s}_h(\vec{r}_{Ni})\vert^2$ depends both on the value of the $s$-like hole wave function $\Psi^{s}_h$ at the position $\vec{r}_{Ni}$ of the atom and on $\beta$, a material dependent quantity describing the coupling of the $p$ electrons in the valence band with the $3d$ electrons on the magnetic atom \cite{Furdyna1988}. For Ni$^{2+}$ in CdTe $\beta$ is expected to be dominated by the antiferromagnetic kinetic exchange \cite{Kacman2001}. Considering that the ground state is a pure hh ($\vert J_h=3/2,J_{hz}=\pm 3/2\rangle$), the exchange interaction acts as an effective magnetic field oriented along the growth axis of the QD. It splits the Ni$^{2+}$ spin states S$_z=\pm$1 resulting into three energy levels associated with S$_z$=0 and S$_z=\pm$1. 
 
For an accurate description of magnetic QDs, a mixing of the ground hh states with higher energy lh states must however be in general taken into account \cite{Tiwari2021}. The last term in (\ref{Hamiltonh}), $\mathcal{H}_{VB}$, describes the energy levels of these mixed hh-lh states. One can first consider the mixing with the lowest energy lh states ($\vert J_h=3/2,J_{hz}=\pm 1/2\rangle$) shifted by an energy $\Delta_{lh}$ depending on the in-plane biaxial strain and the confinement. These states are mixed in the presence of shape or strain anisotropy of the QD. As detailed in appendix A, this can be described by two complex parameters, $P=\delta_{xx,yy}+i\delta_{xy}$ and $Q=\delta_{xz}+i\delta_{yz}$. P stands for the hh-lh mixing induced by an anisotropy in the QD plane (x,y) and Q takes into account an asymmetry in the plane containing the QD growth axis $z$. P is usually written in the form $P=\rho_{vb} e^{-2i\theta_{vb}}$ where $\rho_{vb}$ is the amplitude of the band mixing and $\theta_{vb}$ is the direction of the main anisotropy responsible for the band mixing, measured from the [100] axis.

Considering only an in-plane anisotropy (Q=0), the band mixing couples the hh $J_{hz}=\pm3/2$ and the lh $J_{hz}=\mp1/2$ respectively. Such mixing, allows a flip-flop of the hh spin with another interacting spin. A distortion in a vertical plane (Q$\neq$0) couples the holes $J_z=\pm3/2$ and the $J_z=\pm1/2$ respectively. When exchange coupled with another spin, this term allows a spin-flip of the interacting spin with a conservation of the hh spin (see appendix A).

\section{Model of X$^+$-Ni$^{2+}$ spectra.}

Using the excited state, $\mathcal{H}_{X^+,Ni}$, and the ground state Hamiltonians, $\mathcal{H}_{h,Ni}$, we can calculate the PL spectrum of X$^+$-Ni$^{2+}$. The optical transition probabilities are obtained by calculating the overlap of all the X$^+$-Ni$^{2+}$ and h-Ni$^{2+}$ eigenstates, $\vert\langle h,S_{z}\vert e,S_{z}\rangle\vert^2$. The occupation probabilities of the X$^+$-Ni$^{2+}$ levels are described by an effective spin temperature T$_{eff}$ that is usually larger than the lattice temperature \cite{Kneip2006,Tiwari2020}.

\subsection{Model of QDs with dominant in-plane strain.}

\begin{figure}[hbt]
\centering
\includegraphics[width=0.95\linewidth]{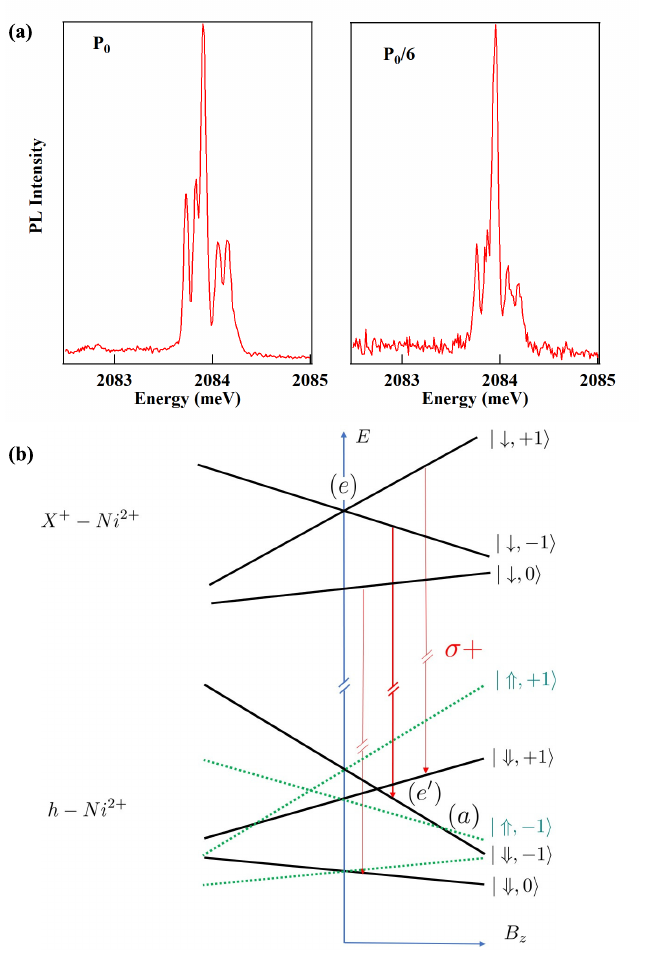}
\caption{(a) Excitation power dependence of the PL of X$^+$-Ni$^{2+}$ in QD1. (b) Scheme of the energy levels involved in the X$^{+}-Ni^{2+}$ $\rightarrow$ h-$Ni^{2+}$ optical transitions. Only the $\sigma$+ transitions, towards a $\Downarrow$ hole, are displayed (solid lines in the h-$Ni^{2+}$ levels).}
\label{Fig:isoscheme}
\end{figure}

For a Ni$^{2+}$, biaxial in-plane strain induces a magnetic anisotropy D$_0$S$_z^2$ oriented along the growth axis of the QDs. The intensity distribution among the 5 lines is controlled by the zero-field splitting D$_0$ and the effective spin temperature which increases with the excitation power \cite{Kneip2006,Tiwari2020}. The emission of QD1 is presented in Fig.\ref{Fig:isoscheme}(a) for two excitation powers. At low power, the contribution of the central line dominates the spectrum. As the excitation power is increased, the contribution of the outer lines increases. This corresponds to a weak and positive value of D$_0$ with the S$_z$=$\pm$1 spin states slightly shifted to high energy.

A slight in-plane strain anisotropy can also mix the S$_z$=$\pm$1 spin states \cite{Tiwari2020JAP}. QD1 (Fig.~\ref{Fig:iso}(b) and Fig.~\ref{Fig:isoB}) corresponds to such local strain configuration. Anti-crossings on the outer lines at zero field (labeled (e) in Fig.~\ref{Fig:isoB}) arise from the mixing, in the excited state, of the Ni$^{2+}$ spin states $\pm1$ by the fine structure term $E$. The same mixing occurs on the h-Ni$^{2+}$ levels when the Zeeman splitting of the $\pm 1$ spin states compensate their exchange interaction with the hole spin (see Fig.~\ref{Fig:isoscheme}(b)). It gives rise to the anti-crossings labeled (e$^{\prime}$) in Fig.~\ref{Fig:isoB}. These anti-crossings correspond to mixing of the Ni$^{2+}$ spin states which does not affect the hole spin and are at the origin of the cross-like behavior observed under B$_z$ in $\sigma$+ polarization.

\begin{table}[!hbt]
    \center
\begin{tabular}{c|ccccccccccccc}
  Ni$^{2+}$          & I$_{hNi}$ & I$_{eNi}$ & D$_{0}$ & E &  g$_{Ni}$ & \\
            & $\mu$eV   & $\mu$eV   & meV     & meV &  \\
            \hline 
  & 110  & -10 & 0.5 & -0.1 & 2.0 & \\
            \hline \hline
        X$^+$       & $\Delta_{lh}$ & $\rho_{vb}$ & $\theta_{vb}$ & $\gamma$  & g$_e$  & g$_h$  & $\delta_{xz}$ & $\delta_{yz}$ &\\
            &        meV    &     meV  &  $^\circ$  & $\mu$eV$T^{-2}$   &   & &   meV  &    meV   \\
            \hline
    & 25 &  4 &  45  &   2   & -0.2  &  0.7 & 2 & 0  \\  
\end{tabular}
    \caption{Parameters used in the modeling of X$^+$-Ni$^{2+}$ in Fig.~\ref{Fig:modBiso}(a) and (c).}
    \label{TableParXciso}
\end{table}

\begin{figure*}[hbt]
\centering
\includegraphics[width=1\linewidth]{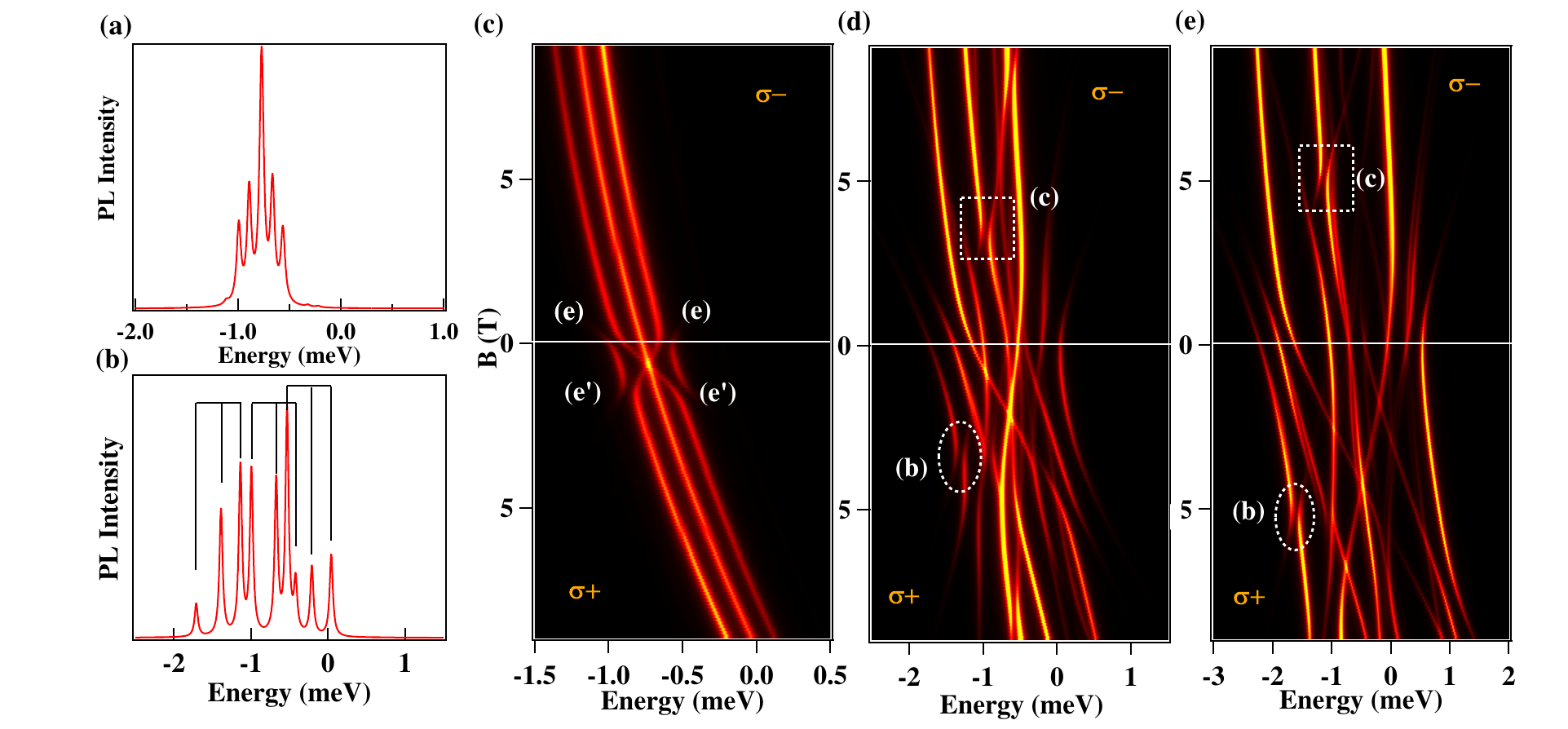}
\caption{(a) and (b) calculated PL spectra with the parameters of table \ref{TableParXciso} and table \ref{TableParXcisob} respectively. (c) (d) and (e): Magnetic field dependence of the PL of X$^+$-Ni$^{2+}$ calculated with the parameters of: (c) table \ref{TableParXciso}, (d) table \ref{TableParXcisob} and (e), table \ref{TableParXcisob} with I$_{hNi}$=700$\mu$eV and I$_{eNi}$=-100$\mu$eV. Lines are broadened with a Lorentzian of FWHM of 50 $\mu$eV. A thermalization on the X$^+$-Ni$^{2+}$ levels with an effective spin temperature T$_{eff}$=20 K is used. In all models, zero energy corresponds to the energy of hh X$^+$ without exchange interaction and lh-hh mixing.}
\label{Fig:modBiso}
\end{figure*}

Spectra of X$^+$-Ni$^{2+}$ calculated with the parameters listed in table \ref{TableParXciso} are presented in Fig.\ref{Fig:modBiso}. The weak overall splitting of QD1 around 330 $\mu$eV is obtained with an anti-ferromagnetic hole-Ni$^{2+}$ exchange interaction I$_{hNi}$=110 $\mu$eV. This expected anti-ferromagnetic coupling \cite{Kacman2001} is confirmed by the intensity distribution under large magnetic field. The Ni$^{2+}$ spin is split by g$_{Ni}\approx$2 and under a large B$_z$ it thermalises on the lowest spin states S$_z$=-1. A $\sigma +$ recombination for instance live in the dot a spin down hole ($\Downarrow$) and the largest intensity line corresponds to a recombination toward $\vert\Downarrow,-1\rangle$ (Fig.\ref{Fig:isoscheme}). As observed in Fig.~\ref{Fig:isoB}, in $\sigma+$ polarization most the PL intensity comes from the lower energy line showing that $\vert\Downarrow,-1\rangle$ is shifted to high energy by an anti-ferromagnetic hole-Ni$^{2+}$ exchange interaction. 

For a general description, a hh-lh mixing induced by in-plane strain anisotropy and shear strain are included in the model. The model reproduces the main feature of the observed spectra of QD1 with in particular the splitting at zero field of the outer lines and the characteristic cross-like behavior under magnetic field in $\sigma$+ polarization which are controlled by a local anisotropy term E=-0.1 meV. An e-$Ni^{2+}$ exchange interaction much lower than the emission linewidth (I$_{eNi}$=-10 $\mu eV$, ten times smaller than I$_{hNi}$) is used in this model. A larger value would introduce additional anti-crossings due to the e-Ni$^{2+}$ flip-flops \cite{Tiwari2021} which are not observed in the experiment. 

However, the model fails to explain the large anti-crossings, labeled (a) and (a$^{\prime}$) in Fig.~\ref{Fig:isoB}, observed on the outer lines in each circular polarisation around B$_z$=1.5 T. These anti-crossings occur when the low energy line in $\sigma$+ polarization (corresponding to the final state $\vert \Downarrow,-1\rangle$) overlaps with the high energy line in $\sigma$- polarization (corresponding to the final state $\vert \Uparrow,-1\rangle$). This happens, under a positive magnetic field, when the h-Ni$^{2+}$ states $\vert \Downarrow,-1\rangle$ and $\vert \Uparrow,-1\rangle$ overlap (see level scheme in Fig.~\ref{Fig:isoscheme}). The corresponding magnetic field is controlled by the exchanged induced splitting of the Ni$^{2+}$ spin at zero field and by $g_h$. The observed anti-crossing would correspond to a spin-flip of the hole which preserves the spin of the Ni$^{2+}$. Such spin-flips are not permitted in the presented model and will be discussed in Section V.

\subsection{Influence of strain disorientation on the X$^+$-Ni$^{2+}$ spectra.}

\begin{figure}[hbt]
\centering
\includegraphics[width=0.9\linewidth]{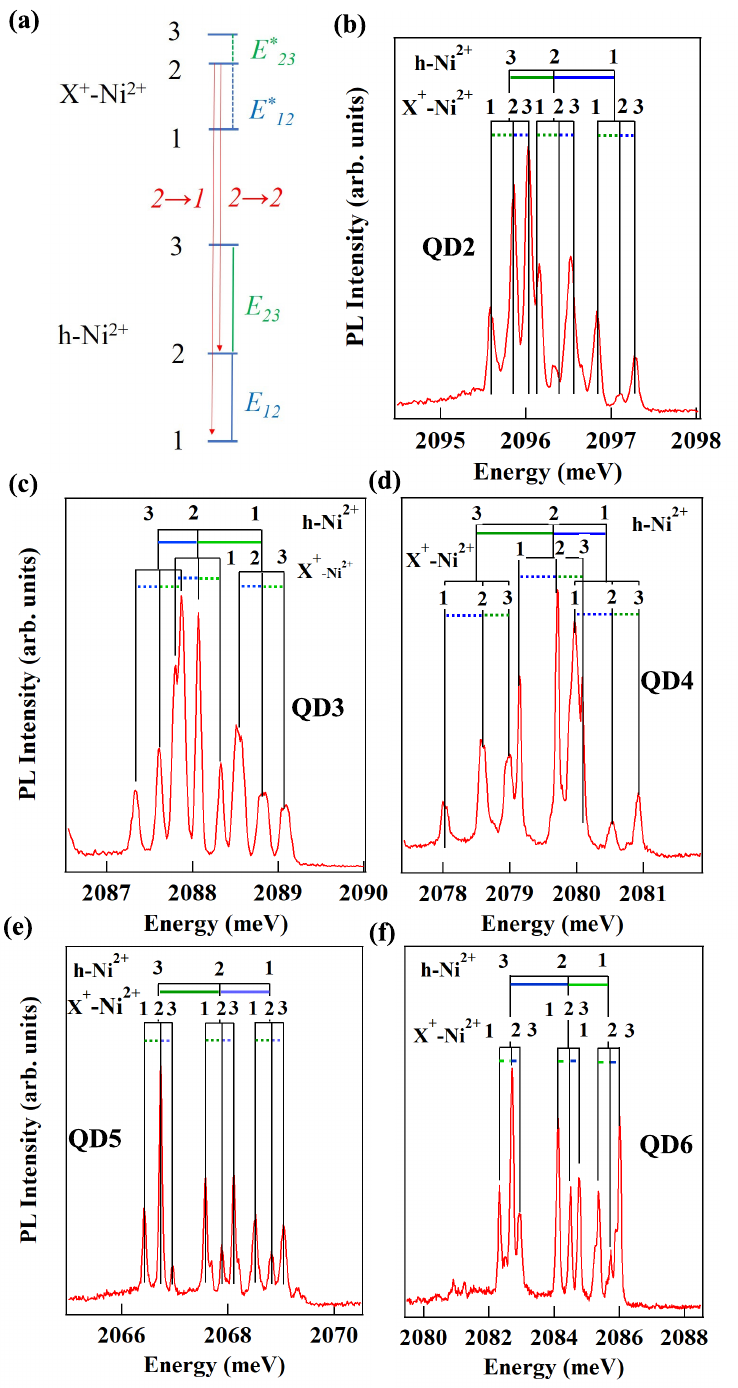}
\caption{(a) Scheme of the energy levels in the ground (h-Ni$^{2+}$) and in the excited (X$^+$-Ni$^{2+}$) states of charged and Ni$^{2+}$-doped QDs with strain disorientation. (b), (c), (d), (e) and (f), detailed zero field spectra of X$^+$-Ni$^{2+}$ in the investigated QDs. For each optical transition, the corresponding energy levels in the excited and in the ground states are identified by their number.}
\label{Fig:Levels}
\end{figure}

The presence of a disorientation of the strain at the magnetic atom location is at the origin of the more complex spectra observed for X$^+$-Ni$^{2+}$ in most of the dots \cite{Krebs2009}. The energy level structure of X$^+$-Ni$^{2+}$ is controlled by $\mathcal{H}_{Ni}$ and by the electron-Ni$^{2+}$ exchange. For a disoriented strain frame, the diagonalization of $\mathcal{H}_{Ni}$ results in 3 levels corresponding to mixed S$_z$ spin states. A scheme of the resulting general level structure in a charged dot is presented in Fig.\ref{Fig:Levels}(a). The electron-Ni$^{2+}$ interaction in strained QDs is expected to be much weaker than the zero field splitting of the atom and the 3-levels structure of the Ni$^{2+}$ is not significantly affected by this interaction. In the ground state, the 3 levels are further split by the exchange interaction with the hole spin which acts as an effective magnetic field aligned along the growth axis of the QD. This leads to eigenstates with Ni$^{2+}$ angular momentum very different in the initial and final state so that all optical transitions become partially allowed. Up to nine emission lines, labeled $i\rightarrow f$ (with i,f=1,2,3), can be obtained \cite{Krebs2009}. For example, the highest and the lowest energy PL lines correspond to the transitions $3 \rightarrow 1$ and $1 \rightarrow 3$ respectively. The order of appearance of the other transitions in the PL spectra depends on the details of the energy splittings in the excited and the ground states.

\subsubsection{Energy level structure deduced from the PL spectra.}

A careful analysis of the experimental PL spectra shows that this 3 x 3 energy level structure is directly observed in the energy spacing of the emission lines. This is illustrated in Fig.~\ref{Fig:Levels} and is particularly clear for the large splitting dots (QD5 and QD6) where the spectra can be decomposed into three 3-lines groups. A similar structure can be deduced from the spectra of dots with a lower overall splitting (QD1, QD3 and QD4 in Fig.~\ref{Fig:Levels}). In this case, the 3-lines groups slightly overlap.  

It follows from the level scheme of Fig.~\ref{Fig:Levels}(a) that the energy spacing within the 3-lines groups is controlled by the structure of the excited state (X$^{+}$-Ni$^{2+}$) which has the smallest splitting. The energy separation between the 3-lines groups is determined by the splitting of the h-Ni$^{2+}$ complex in the ground state. In Fig.~\ref{Fig:Levels}, each PL line in each QD is labeled with its initial and final state. The energy splitting in the ground (E$_{ij}$) and in the excited states (E$^{\star}_{ij}$) are also displayed with horizontal bars. Each two-levels energy splittings appears three times in the emission spectra \cite{Krebs2009}.

\begin{table}[!hbt]
    \center
\begin{tabular}{c|ccc|ccc}
& &h-Ni$^{2+}$ & & &  X$^+$-Ni$^{2+}$  \\

& $E_{12}$ & $E_{23}$ & $E_{13}$ & $E^{\star}_{12}$ &  $E^{\star}_{23}$ & $E^{\star}_{13}$ \\
\hline \hline
QD2         & 0.75  & 0.50  & 1.25 & 0.26 & 0.17  & 0.43    \\
QD3			& 0.75  & 0.46  & 1.21 & 0.28 & 0.26  &  0.54   \\
QD4         & 0.83  & 1.12  & 1.95 & 0.56 & 0.38  &  0.94   \\
QD5         & 0.95  & 1.15  & 2.1  & 0.3  & 0.22  &  0.52   \\
QD6			& 1.25  & 1.79  & 3.04 & 0.39 & 0.25  &  0.64   \\
\hline \hline   
\end{tabular}
\caption{Energy splittings in meV extracted from the PL spectra of the investigated QDs. E$_{ij}$ for the ground state, E$^*_{ij}$ for the excited state.}
\label{TableSplit}
\end{table}

As all the optical transitions are allowed, some of them share the same excited state and form $\Lambda$ type optical systems (see an example in Fig.~\ref{Fig:Levels}(a)). An excitation on one of the branch of the $\Lambda$ system can produce some PL on the other branch. The successive excitation of these $\Lambda$ systems is at the origin of the resonances observed in the resonant-PL spectra. For example, a resonant excitation on the transition 1$\rightarrow$1 produces an X$^+$-Ni$^{2+}$ on the low energy level (1). This can give rise to a resonant PL on the lower energy lines 1$\rightarrow$2 and 1$\rightarrow$3. This is clearly observed for QD3 in Fig.~\ref{Fig:PLE}(a) where an excitation on line (7) (1$\rightarrow$1) gives rise to a PL on the lower energy lines (1) (1$\rightarrow$3) and (3) (1$\rightarrow$2). A similar behavior is observed in the resonant PL for all the investigated charged dots (see Fig.~\ref{Fig:PLE}).  

The identification of each optical transition permits the energy splittings in the ground and excited states of the charged Ni$^{2+}$-doped QDs to be accurately determined. The measured splittings are listed in table \ref{TableSplit}. The overall splitting in the excited states E$^*_{13}$, only slightly perturbed by the electron-Ni$^{2+}$ interaction remains lower than 1 meV in all the investigated QDs. This gives an order of magnitude of the magnetic anisotropy term D$0$. The spin splittings are significantly larger in the ground state (h-Ni$^{2+}$). This results from the exchange interaction with the hole spin which increases the splitting of Ni$^{2+}$ spin levels.

\subsubsection{Magnetic field dependence of the PL of X$^{+}$-Ni$^{2+}$.}

This level structure is confirmed by comparing the experimental and calculated magnetic field dependence of X$^{+}$-Ni$^{2+}$ emission. A spectra calculated at zero field with the parameters of table \ref{TableParXcisob} is presented in Fig.\ref{Fig:modBiso}(b). The emission organized into three 3-lines groups at zero field can be clearly reproduced. The  most important parameter which controls the appearance of the 9 optical transitions is $\theta_{s}$, the angle between the $z$ axis and z$^{\prime\prime}$, the main axis of the disoriented strain frame.

\begin{table}[!hbt]
    \center
\begin{tabular}{c|ccccccccccccc}
 Ni$^{2+}$            & I$_{hNi}$ & I$_{eNi}$ & D$_{0}$ & E &  g$_{Ni}$ &  $\theta_{s}$ & $\varphi_{s}$ & $\psi_{s}$\\
            & $\mu$eV   & $\mu$eV   & meV     & meV &  &  $^\circ$ & $^\circ$ & $^\circ$ \\
            \hline 
 & 350  & -50 & 0.4 & -0.1 & 2.0 & 40 & -30 & 0 \\
            \hline \hline
      X$^+$        & $\Delta_{lh}$ & $\rho_{vb}$ & $\theta_{vb}$ & $\gamma$     & g$_e$  & g$_h$  & $\delta_{xz}$ & $\delta_{yz}$ \\
            &        meV    &     meV  &  $^\circ$  & $\mu$eV$T^{-2}$   &   & &   meV  &    meV      \\
            \hline 
     & 25 &   4 &  45  &   2   & -0.2  &  0.7 & 2 & 0 \\
            
\end{tabular}
    \caption{Parameters used in the modeling of the charged exciton of Fig.~\ref{Fig:modBiso}(d).}
    \label{TableParXcisob}
\end{table}

In the model, the splitting within the 3-lines group is controlled by the Ni$^{2+}$ fine structure parameters and the orientation of the strain ({\it i.e.} the fine structure of e-Ni$^{2+}$). The larger splitting between the 3-lines groups depends on the energy levels structure of h-Ni$^{2+}$. It is controlled by the exchange interaction with the confined hole and the 3-lines groups can possibly overlap for a weak I$_{hNi}$. 

Fig.~\ref{Fig:modBiso}(d) presents a calculated magnetic field dependence of the PL spectra. The main feature of the experimental spectra can be explained by the local strain disorientation. The chosen h-Ni$^{2+}$ exchange interaction in Fig.~\ref{Fig:modBiso}(d) corresponds to the general case of dots like QD3 where the 3-lines groups overlap. An increase of I$_{hNi}$ increases the overall splitting of the PL spectra and separates the 3-lines groups (see Fig.~\ref{Fig:modBiso}(e)).

In these calculations, anti-crossings labeled (b) and (c) are observed in the magnetic field dependence when a valence band mixing induced by in-plane anisotropy is included (P$\neq$0). Such anti-crossings are observed in most of the experimental spectra (see Fig.~\ref{Fig:mapB}) and are usually larger than the calculated values. In the model, the value of these anti-crossings are controlled by the amplitude of the band mixing and by the angles $\theta_{s}$, $\varphi_{s}$ and $\psi_{s}$. The magnetic field position of these anti-crossings and their evolution with the change of the overall zero field splitting of the PL spectra is also well reproduced by the model (see Fig.~\ref{Fig:modBiso}(d) and (e)). The model also accurately reproduces the magnetic field positions of these anti-crossings and, as illustrated in Fig.~\ref{Fig:modBiso} (d) and (e), how these positions evolve with changes of the overall zero-field splitting of the PL spectra.

\subsubsection{Linear polarization properties of X$^+$-Ni$^{2+}$}

\begin{figure}[!hbt]
\centering
\includegraphics[width=1\linewidth]{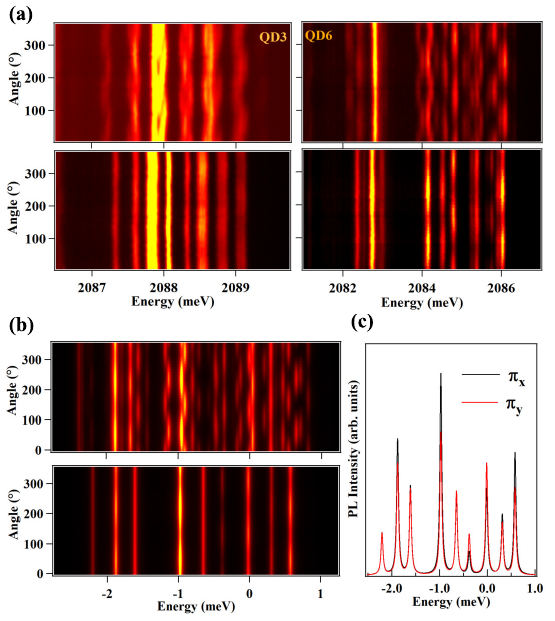}
\caption{(a) Linear polarization properties of the PL of X$^+$-Ni$^{2+}$ in QD3 and QD6. Top: Linear polarization under B$_x$=2T. Bottom: Linear polarization at B=0T. (b) Linear polarization properties of X$^+$-Ni$^{2+}$ calculated with the parameters of table \ref{TableParXcisob} and I$_{hNi}$=700 $\mu$eV, I$_{eNi}$=-100$\mu$eV, $\theta_{Bx}$=0. Top panel: Linear polarization dependence of the PL intensity map at B$_x$=2T. Bottom panel: Linear polarization dependence of the PL intensity map at B=0T. (c) Calculated linearly polarized PL spectra at B=0T.}
\label{Fig:Bx}
\end{figure}

A linear polarization rate can be observed in some charged excitons (see Fig.~\ref{Fig:NiCPL} and Fig.~\ref{Fig:Bx}). When the 3-lines groups are well separated, the linear polarization is more pronounced on the two high energy groups of lines (see QD5 and QD6). The lowest energy triplet shows weaker linear polarization. Within the two high energy 3-lines groups, the high and the low energy lines have orthogonal linear polarization directions and two groups present a similar linear polarization structure but with a $\pi/2$ phase shift between them. 

A magnetic field applied in the plane of the dots, B$_x$, also induces a linear polarization. A transverse field dependence of X$^{+}$-Ni$^{2+}$ is presented in Fig.~\ref{Fig:Bx} for two QDs with very different overall zero field splitting, QD3 and QD6. B$_x$ induces a Zeeman energy which splits linearly polarized lines. For QD6, with well separated 3-lines groups, the linear polarization is more pronounced on the two high energy 3-lines groups as observed for the linear polarization at zero magnetic field. 

The developed model also qualitatively explains the general behavior of the linear polarization structure at zero field and under a transverse magnetic. In the calculated linearly polarized PL intensity maps presented in Fig.~\ref{Fig:Bx}, a more pronounced polarization is obtained on the two high energy group of lines both at zero-field and under a transverse magnetic field. Despite a discrepancy in intensities, the sequence of polarised lines at zero field is also reproduced. Orthogonal linear polarization directions are observed for the high- and low-energy lines of the 3-line groups, as well as a $\pi/2$ shift in the polarization sequence for the two high-energy groups.

To qualitatively understand these polarization properties, let us recall that, as shown in Mn-based magnetic QDs \cite{Varghese2014,Leger2006}, the linear polarization on the charged exciton at zero field results from the spin-flip interaction between the magnetic atom and the hole induced by the presence of the hh-lh mixing. The lowest energy 3-lines group, which is the less linearly polarized, corresponds to recombination towards the high energy h-Ni$^{2+}$ state (3). Because of the antiferromagnetic hole-Ni$^{2+}$ exchange interaction, this high energy level corresponds to mainly parallel hole and Ni$^{2+}$ spins. The hole-Ni$^{2+}$ flip-flops induced by hh-lh mixing are then significantly blocked within this level, preventing the occurrence of significant linear polarization. This level is also apparently less sensitive to a transverse magnetic field.

\subsection{Influence of the mixing of hole subbands.}

In the event of disorientation of the strain frame, hh-lh mixing induced by the shape or strain anisotropy of the dots can explain the anti-crossings observed in the magnetic field dependence of the PL. The value of the anti-crossings depends on the value of the mixing but also on the disorientation angles. For dots like QD1 (Fig.~\ref{Fig:isoB}), where the strain frame remains oriented along the QD growth axis, anti-crossing (a) and (a$^{\prime}$) are not explained by the presence of the hh-lh mixing discussed until now.

Other sources of valence band mixing can be considered in magnetic dots to explain details in their emission spectra. It was in particular demonstrated by Bhattacharjee {\it et al.} that, in spherical nanocrystals, for a magnetic atom not located in the center of the dot, the hole-Ni$^{2+}$ exchange interaction can be influenced by the mixing of the hh ground state with higher energy hole states \cite{Bhatta2007}. 

When high energy hole states are taken into account, non-diagonal terms of the Kohn-Luttinger Hamiltonian $\mathcal{H}_{KL}$ (see Appendix B) induces an inter-subband mixing even in the absence of anisotropy of the dot \cite{Pedersen1996,Broido1992,Chutia2008}. In the simplest case of a spherical nanocrystal considered by Bhattacharjee {\it et al.}, $\mathcal{H}_{KL}$ introduces a $d$-orbital component in the envelope function of the ground hh state \cite{Bhatta2003,Bhatta2004}. Since the hole/magnetic atom exchange interaction is a probe of the hole envelope function at the magnetic atom site, this introduces a position-dependent part into the exchange interaction. The exchange interaction must be written as a spin Hamiltonian with the isotropic part (\ref{ExIso}) and an additional anisotropic part when the magnetic atom is not located at the center of the QD \cite{Bhatta2007,Krebs2009}.

\subsubsection{Mixing of hole states in a lens-shape quantum dot.}

The model of Bhattacharjee {\it et al.} \cite{Bhatta2007,Krebs2009} based on a spin-effective Hamiltonian with cylindrical symmetry is not completely suitable for magnetic self-assembled QDs as (i) hh/lh mixing is already present in non-magnetic self-assembled dots and (ii) the symmetry of a dot with an off-center magnetic atom is much lower than for a nano-crystal. A more realistic model of confinement is a lens-shaped dot with a parabolic potential in the plane ($V_{\parallel}(\rho)=\frac{1}{2}K\rho^2$ where K measures the strength of the potential) and a finite square well along the growth direction ($V_{\bot}(z)=\Delta E_v$ for $\vert z\vert\geq w/2$ and $V_{\bot}(z)=0$ for $\vert z\vert< w/2$ where $w$ is the width of the quantum well) \cite{Pedersen1996}. The use of $\mathcal{H}_{KL}$ within such confinement potential geometry allows to deduce coupling terms between hole levels and their consequences for the hole-Ni$^{2+}$ exchange interaction. 

In this confinement potential, a general form for a hole wave function is:

\begin{eqnarray}
\psi(\vec{r})=\sum_{j_z}F_{j_z}(\vec{r})\vert\frac{3}{2},j_z\rangle
\label{wave}
\end{eqnarray}

\noindent where $\vert\frac{3}{2},j_z\rangle$ is the band edge Bloch function and F$_{j_z}$(r) the envelope function. Since the potential has cylindrical symmetry the envelope has a definite angular momentum. Following F.B. Pedersen {\it et al.} \cite{Pedersen1996} we can define the total angular momentum F=J+L where J is the angular momentum of the Bloch function and L the envelope angular momentum. In the axial approximation of $\mathcal{H}_{KL}$ (see Appendix B), F$_z$ is a constant of the motion \cite{Pedersen1996}. It is possible to find the eigenstates of $\mathcal{H}_{KL}$ and F$_z$ simultaneously. The eigenstates can be labeled with the quantum number f$_z$ and a hole state be written in cylindrical coordinates as:

\begin{eqnarray}
\psi(\vec{r})=\sum_{j_z}F_{j_z}(\rho,z)e^{i\varphi(f_z-j_z)}\vert\frac{3}{2},j_z\rangle
\label{wave}
\end{eqnarray}

If the band mixing is neglected (i.e. neglect non-diagonal terms of $\mathcal{H}_{KL}$), all the lh and hh levels are decoupled. The envelope function can be decomposed into an in-plane and a subband part:  

\begin{eqnarray}
\phi_{n_r,l,s}(\vec{r})=\Phi_{n_r,l}(\rho,\varphi)f_s(z)
\label{wave}
\end{eqnarray}

\noindent where $f_s(z)$ is the wave function of the subband $s$ of the quantum well.  $\Phi_{n_r,l}(\rho,\varphi)$ is the 2D harmonic oscillator wave function (see Appendix C for more details) describing the in-plane motion with n$_r$=(n-$\vert l\vert$)/2 the radial quantum number, $n$=0, 1, ... the principal quantum number and $l$=-n,-n+2,...,n-2,n the azimuthal quantum number \cite{Jacak}.

With this appropriate basis, the hole wave function in a lens-shape QD for states with total angular momentum $f_z=j_z+l$ can be expanded as: 

\begin{eqnarray}
\psi_{f_z}(\vec{r})=\sum_{j_z,n_r,s}C(n_r,s,j_z)\phi_{n_r,f_z-j_z,s}(\vec{r})\vert\frac{3}{2},j_z\rangle
\label{wave}
\end{eqnarray}

$\mathcal{H_{KL}}$ mixes states with the same $f_z$ \cite{Pedersen1996}. The lowest energy "s"-like ground heavy-hole of the subband HH1, $\phi_{0,0,1}(\vec{r})\vert\frac{3}{2},+\frac{3}{2}\rangle$ (j$_z$=+3/2, l=0, n$_r$=0, s=1) with $f_z=j_z+l$=3/2 is not a pure hh state. For a qualitative description of the influence of band mixing, we can limit the development to the lowest energy subbands LH2, LH1 and HH2 and to the lowest energy states of each subband. $\mathcal{H_{KL}}$ couples $\phi_{0,0,1}(\vec{r})\vert\frac{3}{2},+\frac{3}{2}\rangle$ with: 
 
- the "{\it p}"-like light-hole in the second subband LH2  $\phi_{0,1,2}(\vec{r})\vert\frac{3}{2},\frac{1}{2}\rangle$ (j$_z$=+1/2, l=1, n$_r$=0, s=2). 

- the "{\it d}"-like light-hole state in the first subband LH1, $\phi_{0,2,1}(\vec{r})\vert\frac{3}{2},-\frac{1}{2}\rangle$ (j$_z$=-1/2, l=2, n$_r$=0, s=1). 

- the "{\it f}"-like heavy-hole state in the second subband HH2 $\phi_{0,3,2}(\vec{r})\vert\frac{3}{2},-\frac{3}{2}\rangle$ (j$_z$=-3/2, l=3, n$_r$=0, s=2).

The wave function of the lowest energy heavy-hole states can then be written:

\begin{eqnarray}
\psi_{+3/2}(\vec{r})=\phi_{0,0,1}(\vec{r})\vert\frac{3}{2},+\frac{3}{2}\rangle+C_1\phi_{0,2,1}(\vec{r})\vert\frac{3}{2},-\frac{1}{2}\rangle\\ \nonumber
+C_2\phi_{0,3,2}(\vec{r})\vert\frac{3}{2},-\frac{3}{2}\rangle+C_3\phi_{0,1,2}(\vec{r})\vert\frac{3}{2},+\frac{1}{2}\rangle
\label{pp}
\end{eqnarray}

\noindent and similarly 

\begin{eqnarray}
\psi_{-3/2}(\vec{r})=\phi_{0,0,1}(\vec{r})\vert\frac{3}{2},-\frac{3}{2}\rangle+C_1\phi_{0,-2,1}(\vec{r})\vert\frac{3}{2},+\frac{1}{2}\rangle\\ \nonumber
+C_2\phi_{0,-3,2}(\vec{r})\vert\frac{3}{2},+\frac{3}{2}\rangle+C_3\phi_{0,-1,2}(\vec{r})\vert\frac{3}{2},-\frac{1}{2}\rangle
\label{mm}
\end{eqnarray}

With these mixed states we can evaluate the structure of the hole-Ni$^{2+}$ exchange interaction (\ref{Hamiltonh}). We obtain, up to the first order in C$_i$ :

\begin{eqnarray}
\langle\pm3/2\vert\mathcal{H}_{h-Ni}\vert\pm3/2\rangle=\pm 3/2I^0_{hNi}S_z+\xi I^0_{hNi} S_{\mp}
\end{eqnarray}

\noindent where the first term corresponds to the isotropic part of the hh-Ni$^{2+}$ exchange interaction with $I^0_{hNi}$ the exchange integral of the pure ground heavy-hole with a magnetic atom located at $\vec{r_a}(\rho_a,\varphi_a,z_a)$ \cite{Bhatta2007}. In the second term, $\xi$ depends on C$_3$ and on the position of the magnetic atom $\vec{r_a}(\rho_a,\varphi_a,z_a)$ through the overlap of the hole envelope functions with the atom, $\phi_{n_r,l,s}(\vec{r}_a)$. This term conserves the hole spin and flips the spin of the magnetic atom (like the Q term in $\mathcal{H}_{VB}$).

It also results from development of the mixed states that:

\begin{eqnarray}
\langle+3/2\vert\mathcal{H}_{h-Ni}\vert-3/2\rangle=\epsilon I^0_{hNi}S_{-}+\eta I^0_{hNi}S_z
\label{KLaniso}
\end{eqnarray}

\noindent where $\epsilon$ depends on C$_1$ and $\eta$ depends on C$_2$. The first term in (\ref{KLaniso}) corresponds to a hh/Ni$^{2+}$ spin flip-flop and acts as the P term in a QD with an in-plane anisotropy. The second term, controlled by $\eta$, corresponds to a spin-flip of the hh conserving the Ni$^{2+}$ spin. This additional term has no counterpart in an anisotropic QD described by $\mathcal{H}_{VB}$. $\epsilon$ and $\eta$ are related to the position of the magnetic atom $\vec{r_a}$ through the overlap of the hole envelope functions with the atom $\phi_{n_r,l,s}(\vec{r}_a)$. As $\phi_{n_r=0,l\neq0,s}(\vec{0})$=0, the contribution of these additional terms cancel for a magnetic atom located at the center of the dot and can only be significant for a magnetic atom far from the $z$ axis.  

In summary, taking into account $\mathcal{H}_{KL}$ and the lowest energy levels in a lens-shaped QD shows that, in addition to the isotropic Heisenberg type exchange interaction, at least three additional parameters are required to describe the coupling of the hole and Ni$^{2+}$ spins:
 
- A hh-Ni$^{2+}$ flip-flop term dependent on $\epsilon$ (equivalent to the P term in an anisotropic QD described by $\mathcal{H}_{VB}$). 

- A spin flip term that preserves the hole spin and flip the magnetic atom dependent on $\xi$ (equivalent to the Q term in an anisotropic QD described by $\mathcal{H}_{VB}$).

- A spin-flip of the hh preserving the Ni$^{2+}$ spin described by a position dependent parameter $\eta$. This additional term has no counterpart in an anisotropic QD described by $\mathcal{H}_{VB}$.

\subsubsection{Spectra of X$^+$-Ni$^{2+}$ with hole-subbands mixing.}

\begin{figure}[hbt]
\centering
\includegraphics[width=1.0\linewidth]{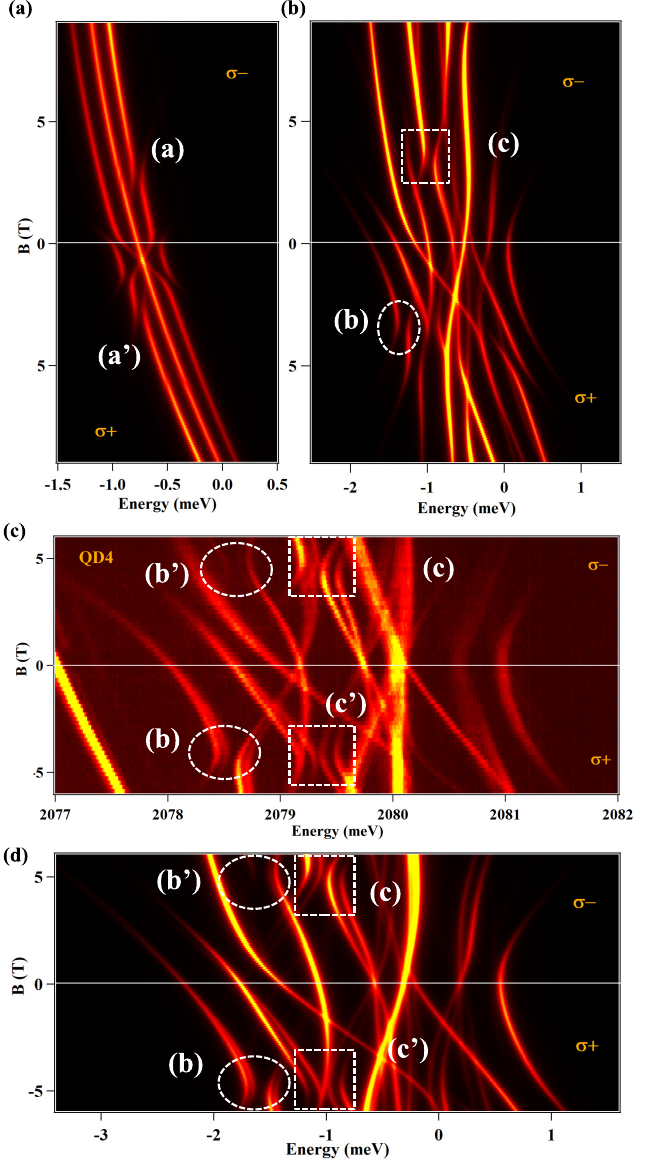}
\caption{Magnetic field dependence of the emission of X$^+$-Ni$^{2+}$ calculated with $\eta$=0.4 and (a) the parameters of table \ref{TableParXciso} and (b) the parameters of table \ref{TableParXcisob}. (c) Detail of the PL at low magnetic field of X$^+$-Ni$^{2+}$ in QD4. (d) Magnetic field dependence of X$^+$-Ni$^{2+}$ calculated with the parameters of table \ref{TableParXcisob} and I$_{eNi}$=-140$\mu$eV, I$_{hNi}$= 575$\mu$eV, D$_0$=0.6 meV, g$_h$=0.8, $\eta$=0.4. An effective temperature T$_{eff}$=20K and a line broadening of 50 $\mu$eV are used for all the calculated spectra.}
\label{Fig:modB}
\end{figure}

For a detailed description of the PL spectra, all the discussed hole/Ni$^{2+}$ spin-flip terms are included in the model. hh/Ni$^{2+}$ flip-flops and flip of the magnetic atom conserving the hh spin can result from a shape or strain anisotropy of the dots or hole-subband mixing by $\mathcal{H}_{KL}$. Both sources of anisotropy are likely to be present in self-assembled QDs and there is no easy way to distinguish between them. Such terms are already included in the model by using $\mathcal{H}_{VB}$. hh spin-flips that preserve the Ni$^{2+}$ spin are only induced by the mixing of hole subbands by the Khon-Luttinger Hamiltonian for an off-center magnetic atom. Results of modeling with an additional spin-flip term proportional to $\eta I_{hNi}S_z$ which couples $J_{hz}=\pm3/2$ and conserve the Ni$^{2+}$ is presented in Fig.~\ref{Fig:modB}. For a dot with a strain frame oriented along the QD growth axis, $\eta$ explains the anticrossing (a) and (a$^{\prime}$) observed in the longitudinal magnetic field dependence (see QD1). The observation of theses large anti-crossings demonstrate the necessity of considering low-symmetry terms in the exchange interaction induced by $\mathcal{H}_{KL}$. This is particularly true for dots with small overall splittings, such as QD1, where the magnetic atom is positioned far from the center, resulting in an increased influence of hole subband mixing. For dots with a disoriented strain frame and a large zero field splitting, this additional term $\eta$ increases the value of the characteristic anti-crossing (b), (b$^{\prime}$) and (c), (c$^{\prime}$) which are already observed in the presence of a P term (Fig.~\ref{Fig:modB}(b)). 

To go deeper into details, we compare in Fig.~\ref{Fig:modB}(c) and Fig.~\ref{Fig:modB}(d),   the emission of QD4 in a low magnetic field with a PL intensity map calculated using this more complete model. Although many of the Ni$^{2+}$ parameters are difficult to adjust independently and achieving a perfect fit is challenging, there is a good level of agreement between the experiment and the model when standard QD parameters are used. In particular, the position and amplitude of the characteristic anticrossings (b) and (c) ((b$^\prime$) and (c$^\prime$) in the opposite polarization) pointed out in all the dots with a sufficiently large splitting in Fig.~\ref{Fig:mapB} are correctly reproduced. 

Anti-crossings labeled (b) and (b$^{\prime}$) occur when the transition $1\rightarrow3$ in $\sigma+$ polarization overlaps the transition $1\rightarrow2$ in $\sigma-$ polarization. Similarly, anti-crossings labeled (c) and (c$^{\prime}$) occur when the $2\rightarrow2$ transition in $\sigma-$ polarization overlaps the $2\rightarrow3$ transition in $\sigma+$ polarization (see Fig.~\ref{Fig:modB}(c)). In both cases, the transitions involved in the anti-crossings share the same mixed Ni$^{2+}$ spin state in the initial state ((1) for (b) and (b$^\prime$), (2) for (c) and (c$^\prime$)) showing that the mixing occurs in the final state. In the final state, both anti-crossings involve the levels ($\Uparrow$,(2)) for the  $\sigma-$ transitions and ($\Downarrow$,(3)) for the $\sigma+$ transitions. These anti-crossings correspond to a hole-Ni$^{2+}$ flip-flop and in the model they can be directly induced by the lh-hh mixing term P responsible of J$_z$-S$_z$ flip-flops. They can also occur in the presence of strain disorientation when the anisotropic part of the exchange interaction, labeled as $\eta$, is included, since, in this case, all the Ni$^{2+}$ S$_z$ spin states are mixed. With such mixing, a flip of the hole retaining S$_z$ state also couple ($\Uparrow$,(2)) and ($\Downarrow$,(3)).

Finally, it can also be noticed that some of the dots, like QD4, present a doubling of the central lines under a weak magnetic field B$_z$. These lines are in particular involved in the anti-crossings (c) and (c$^\prime$). This doubling can also be accurately reproduced by the model (see Fig.~\ref{Fig:modB}(d)). It is due to the weak electron-Ni$^{2+}$ exchange interaction which contributes under B$_z$ to the splitting of the Ni$^{2+}$ spin levels in the initial state, X$^{+}$-Ni$^{2+}$, of the optical transitions. This fine adjustment of the doubling, when observed, permits an independent estimation of the hole-Ni$^{2+}$ and electron-Ni$^{2+}$ exchange interactions. In the modeling, the doubling disappears for I$_{eNi}$=0 and the fact that it cannot be resolved in most dots confirms the dominant contribution of the hole-Ni$^{2+}$ exchange interaction.

\section{Conclusion}

In conclusion, we have demonstrated that magneto-optical micro-spectroscopy can be used to identify the presence of individual Ni$^{2+}$ ions in QDs and extract the main parameters that control their interaction with confined carriers. Our systematic study of numerous dots and our model demonstrate that Ni$^{2+}$-doped QDs exhibit anisotropic strain distribution at the location of the magnetic atom. In the simplest case, an in-plane strain anisotropy mixes the S$_z$=$\pm$ 1 spin states of the atom, causing them to appear as doublets in the emission of charged dots. In most dots, however, the main axis of the strain frame is misaligned with the QD growth axis, resulting in the mixing of all S$_z$ spin states and an increased number of emission lines. A spin-effective model that incorporates strain anisotropy that mixes Ni$^{2+}$ spin states can reproduce the observed PL spectra and their magnetic field dependence.

In charged QDs, the energy splittings resulting from strain disorientation can be accurately deduced from an analysis of the energy spacing in the PL spectra. The energy level structure is also confirmed by resonant PL experiments, which reveal the presence of $\Lambda$-like transitions that can be addressed by resonant optical excitation, giving rise to characteristic resonances in the PL excitation spectra. We have also shown that the presence of hole-Ni$^{2+}$ spin-flip mechanisms, in particular hole spin-flips that conserve the Ni$^{2+}$ spin, are crucial for accurate modeling. These terms originate from the hole-Ni$^{2+}$ exchange interaction combined with lh-hh mixing in anisotropic QDs or hole subband mixing induced by the Kohn-Luttinger Hamiltonian, particularly when the Ni$^{2+}$ ion is off-center in the dot.

\begin{acknowledgements}{}

This work was realized in the framework of the CEA (Institut de Recherche Interdisciplinaire de Grenoble) / CNRS (Institut N\'{e}el) joint research team NanoPhysique et Semi-Conducteurs (NPSC). 

\end{acknowledgements}

\appendix

\section{Influence of strain or shape anisotropy on the hh-lh mixing.}

The mixing of lh and hh induced by shape or strain anisotropy can be described by the Hamiltonian $\mathcal{H}_{VB}$ which is written in the basis ($|\frac{3}{2},+\frac{3}{2}\rangle,|\frac{3}{2},+\frac{1}{2}\rangle,|\frac{3}{2},-\frac{1}{2}\rangle,|\frac{3}{2},-\frac{3}{2}\rangle$) \cite{Chuang, Voon2009}:

\begin{equation}
\mathcal{H}_{VB} = \left(
\begin{array}{cccc}
0                   &Q                               &P                      &0\\
Q^*                 &\Delta_{lh}                     &0                      &P\\
P^*                 &0                               &\Delta_{lh}            &-Q\\
0                   &P^*                             &-Q^*                   &0\\
\end{array}\right)
\label{HBM}
\end{equation}

\noindent where  
\begin{eqnarray}
P=\delta_{xx,yy}+i\delta_{xy}; Q=\delta_{xz}+i\delta_{yz}
\end{eqnarray}

P describes the hh-lh mixing induced by an anisotropy in the QD plane $xy$ and Q takes into account an asymmetry in the plane containing the QD growth axis $z$. The reduction in symmetry can come from the shape of the QD (via the Kohn-Luttinger Hamiltonian) or the strain distribution (via the Bir-Pikus Hamiltonian) \cite{Voon2009}. P is usually written in the form $P=\rho_{vb} e^{-2i\theta_{vb}}$ where $\rho_{vb}$ is the amplitude of the mixing of the lh and hh split by $\Delta_{lh}$ and $\theta_{vb}$ is the direction of the main anisotropy responsible for the band mixing, measured from the [100] axis.

We can qualitatively understand the effect of these mixing terms using limited development. In the presence of anisotropy in the QD plane ($P \neq 0$), the two hole ground states become, in the limit of weak band mixing ($\rho_{vb} \ll \Delta_{lh}$):

\begin{eqnarray}
\vert\Phi^+_h\rangle=\vert+3/2\rangle-\phi\vert-1/2\rangle\nonumber\\
\vert\Phi^-_h\rangle=\vert-3/2\rangle-\phi^*\vert+1/2\rangle
\end{eqnarray}

\noindent with $\phi=\rho_{vb}/\Delta_{lh}e^{2i\theta_{vb}}$ describing the amplitude of the mixing. A first order development of the angular momentum operator $J$ on the subspace of the perturbed holes $\vert\Phi^{\pm}_h\rangle$ leads to 

\begin{eqnarray} \nonumber
\tilde{j}_+= 
\frac{\rho_{vb}}{\Delta_{lh}}\begin{pmatrix}
0 & -2\sqrt{3}e^{-2i\theta_{vb}} \\ 
0 & 0 
\end{pmatrix} \\ 
\tilde{j}_-=
\frac{\rho_{vb}}{\Delta_{lh}}\begin{pmatrix}
0 & 0 \\ 
-2\sqrt{3}e^{2i\theta_{vb}} & 0 
\end{pmatrix}\\ 
\tilde{j}_z=
\begin{pmatrix}
3/2 & 0 \\ 
0 & -3/2 
\end{pmatrix}
\end{eqnarray}

\noindent $\tilde{j}_+$ and $\tilde{j}_-$ flip the hole spin whereas a measurement of the spin projection along $z$ confirms that they are mainly heavy holes. This type of mixing unlock the spin-flips between the hole and its surrounding medium allows a flip-flop of the hh spin with another interacting spin. This is the case when the short-range exchange interaction ($2/3\delta_0(\overrightarrow{\sigma}.\overrightarrow{J})$) couples electron and hole spins allowing an electron-hole flip-flop and a mixing of the two bright excitons. In a magnetic QD the hh/Ni$^{2+}$ flip-flop are also possible.

Similarly, in the presence of distortion in a vertical plane ($Q\neq 0$), the two hole ground states become:

\begin{eqnarray}
\vert\Xi^+_h\rangle=\vert+3/2\rangle+\xi\vert+1/2\rangle\nonumber\\
\vert\Xi^-_h\rangle=\vert-3/2\rangle-\xi^*\vert-1/2\rangle
\end{eqnarray}

A first order development of the angular momentum operator $J$ on the subspace of the perturbed holes $\vert\Xi^{\pm}_h\rangle$ leads to

\begin{eqnarray}\nonumber
\tilde{j}_+= 
\xi\begin{pmatrix}
\sqrt{3} & 0 \\ 
0 & -\sqrt{3} 
\end{pmatrix}; 
\tilde{j}_-=
\xi^*\begin{pmatrix}
\sqrt{3} & 0 \\ 
0 & -\sqrt{3} 
\end{pmatrix}; \\
\tilde{j}_z=
\begin{pmatrix}
3/2 & 0 \\ 
0 & -3/2 
\end{pmatrix}
\end{eqnarray}

Because of this valence band mixing the hole-Ni$^{2+}$ exchange interaction couples the states $\vert\Xi^{\pm}_h,S_z\rangle$ with the states $\vert\Xi^{\pm}_h,S_z+1\rangle$ and $\vert\Xi^{\pm}_h,S_z-1\rangle$. This term corresponds to a spin-flip of the Ni$^{2+}$ spin with a conservation of the hh spin. When combined with the short-range e-h exchange interaction, such band mixing couples $|+1\rangle$ and $|+2\rangle$ excitons on one side and $|-1\rangle$ and $|-2\rangle$ excitons on the other side \cite{Tiwari2021}.

\section{Kohn-Luttinger Hamiltonian in a QD.}

The Khon-Luttinger Hamiltonian describing the hole kinetic energy in the effective mass approximation in the basis ($|\frac{3}{2},+\frac{3}{2}\rangle,|\frac{3}{2},+\frac{1}{2}\rangle,|\frac{3}{2},-\frac{1}{2}\rangle,|\frac{3}{2},-\frac{3}{2}\rangle$) is \cite{Chuang, Voon2009}: 

\begin{equation}
\mathcal{H}_{KL} = \frac{\hbar^2}{2m_0}\left(
\begin{array}{cccc}
H_{hh}                          &R                                      &S                              &0\\
R^*                             &H_{lh}                                 &0                              &S\\
S^*                             &0                                      &H_{lh}                         &-R\\
0                               &S^*                                    &-R^*                           &H_{hh}\\
\end{array}\right)
\label{KL}
\end{equation}

\noindent where

\begin{eqnarray}
H_{hh}=(\gamma_1+\gamma_2)(k_x^2+k_y^2)+(\gamma_1-2\gamma_2)k_z^2
\end{eqnarray}

\begin{eqnarray}
H_{lh}=(\gamma_1-\gamma_2)(k_x^2+k_y^2)+(\gamma_1+2\gamma_2)k_z^2
\end{eqnarray}

\begin{eqnarray}
R=2\sqrt{3}\gamma_3ik_-k_z
\end{eqnarray}

\begin{eqnarray}
S=\sqrt{3}\gamma k_-^2+\sqrt{3}\gamma^{\prime} k_+^2
\end{eqnarray}

\noindent with 

\begin{eqnarray}
\gamma=(\gamma_1+\gamma_2)/2, \gamma^{\prime}=(\gamma_2-\gamma_3)/2
\end{eqnarray}

\noindent and

\begin{eqnarray}
k=-i\nabla; k_{\pm}=k_x\pm ik_y
\end{eqnarray}

Here, $\gamma_1$, $\gamma_2$ and $\gamma_3$ are the Luttinger parameters \cite{Voon2009}. This Hamiltonian can be further simplified by adopting the axial approximation to preserve cylindrical symmetry about the z-axis. It consists in setting $\gamma^{\prime}=0$. This approximation is sufficient to qualitatively described holes subbands mixing in lens-shape QDs \cite{Chutia2008}.

\section{2D harmonic oscillator wave-function.}

The wave function of a two-dimensional harmonic oscillator can be written:

\begin{eqnarray}
\Phi_{n_r,l}(\rho,\varphi)=\\ \nonumber 
\frac{e^{il\varphi}}{\sqrt{2\pi}}\frac{\sqrt{2}}{a_h}\sqrt{\frac{n_r!}{(n_r+\vert l\vert)!}}\left(\frac{\rho}{a_h}\right)^{\vert l\vert} e^{-\frac{\rho^2}{2a_h^2}}\mathcal{L}_{n_r}^{(\vert l\vert)}\left(\frac{\rho^2}{a_h^2}\right)
\label{wave}
\end{eqnarray}

\noindent where $\mathcal{L}_{n_r}^{(\vert l\vert)}\left(z\right)$ denotes the generalized Laguerre polynomials \cite{Jacak}. This wave-function depends on $n_r$=(n-$\vert l \vert$)/2 the radial quantum number with $n$=0,1,... the principal quantum number. $l$ is the azimuthal quantum number with l=-n,-n+2,...,n-2,n. The corresponding eigenenergies are E(n)=$\hbar\Omega$(n+1) with $\Omega=\sqrt{K/m^*}$, where $m^*$ is the effective mass of the confined particle. $a_h$, the characteristic length of the harmonic confinement potential, is given by $a_h^2=\hbar\Omega/K$ \cite{Pedersen1996}.

Together with the wave function of the lowest energy quantum well levels $f_1(z)=\sqrt{\frac{2}{w}} cos(\pi z/ w)$, this wave function permits for instance to calculate the exchange integral of the a pure ground heavy-hole with a magnetic atom located at $\vec{r_a}(\rho_a,\varphi_a,z_a)$:

\begin{eqnarray}
I_h^0=\frac{2}{\pi w a_h^2}\frac{\beta}{3}cos^2(\frac{\pi z_a}{w})e^{-\frac{\rho_a^2}{a_h^2}}
\end{eqnarray}

\end{document}